\documentclass[preprint,aps,prc,showpacs,nofootinbib,secnumarabic,floatfix]{revtex4-1}

\usepackage{graphicx}
\usepackage{bm}
\usepackage{color}
\usepackage{gensymb}
\usepackage{amsmath}
\usepackage{slashed}
\usepackage{appendix}
\usepackage{hyperref}

\newcommand{\gtsim}{\protect\raisebox{-0.5ex}{$\:\stackrel{\textstyle >}
	{\sim}\:$}}

\newcommand{\bvec}[1]{\ensuremath{\boldsymbol{#1}}}

\begin{document}

\title{Coherent interactions of a fast proton with a short-range $NN$ correlation in the nucleus}  

\author{A.B. Larionov$^1$\footnote{e-mail: larionov@theor.jinr.ru},
        Yu.N. Uzikov$^{2,3,4}$\footnote{e-mail: uzikov@jinr.ru}}
  
\affiliation{$^1$ Bogoliubov Laboratory of Theoretical Physics, Joint Institute for Nuclear Research, 141980 Dubna, Russia\\
             $^2$ Laboratory of Nuclear Problems, Joint Institute for Nuclear Research, 141980 Dubna, Russia\\
             $^3$ Department of Physics, Moscow State University, 119991 Moscow, Russia\\
             $^4$ Dubna State University, 141980 Dubna, Russia}

\begin{abstract}
  Nuclear structure at short $NN$-distances is still poorly understood. In particular, the full quantum structure
  of the nucleus with a correlated $NN$-pair is a challenge to theory.
  So far, model descriptions have been limited to the average mean-field picture of the remaining nuclear system after removing the $NN$-pair.
  In the recent experiment of the BM@N Collaboration at JINR \cite{Patsyuk:2021fju},
  the reactions $^{12}\mbox{C}(p,2pn_s)^{10}\mbox{B}$ and $^{12}\mbox{C}(p,2pp_s)^{10}\mbox{Be}$
  induced by the hard elastic $pp$ scattering were studied.
  Here, $n_s$ or $p_s$ denote the undetected slow nucleon in the rest frame of $^{12}\mbox{C}$.    
  In contrast to the previous experiments, the residual bound nucleus was also detected
  which requires a new level of theoretical understanding.
  In the present work, we apply the technique of fractional parentage coefficients of the translationally-invariant
  shell model (TISM) to calculate the spectroscopic amplitude of the system $NN-B$ where $B$ is the remaining nuclear system.
  The spectroscopic amplitude enters the full amplitude of a nuclear reaction. 
  The relative $NN-B$ wave function is no longer a free parameter of the model
  but is uniquely related to the internal state of $B$.
  The interaction of the target proton with the $NN$-pair is considered in the impulse approximation.
  We also include the initial- and final state interactions of absorptive type as well as the single charge exchange processes.
  Our calculations are in a reasonable agreement with the BM@N data.
\end{abstract}

\maketitle

\section{Introduction}
\label{intro}

Short-range $NN$ correlations (SRCs) in nuclei are in the focus of experimental and theoretical studies since about three decades,
see recent reviews in Refs.~\cite{CiofidegliAtti:2015lcu,Hen:2016kwk}. It is nowadays well established that in medium-to-heavy nuclei
about 20-25\% of nucleons are in the state of SRCs.
These nucleons populate the part of the nucleon momentum distribution above Fermi momentum of $\sim 250$ MeV/c \cite{Piasetzky:2006ai}.
Most SRCs are the $pn$ ones, although the fraction of $pp$-SRCs increases with
missing momentum
\footnote{The missing momentum $p_{\rm miss}$
is defined as the momentum of the struck proton in the nucleus rest frame before knock-out.}
at $p_{\rm miss}=400-800$ MeV/c indicating
the transition from dominating tensor- to dominating repulsive scalar interaction,
which follows from the analysis of the reaction
$^4\mbox{He}(e,e^\prime pN)$ at JLab \cite{LabHallA:2014wqo}.
This conclusion is also supported by the increasing cross-section ratio $A(e,e^\prime pp)/A(e,e^\prime p)$ with $p_{\rm miss}$ at $p_{\rm miss}=400-600$ MeV/c,
independent on the target nucleus \cite{CLAS:2020mom}.

Most experimental searches for SRCs have been carried out by detecting a scattered particle ($e$ or $p$), a recoil proton and its partner nucleon,
c.f. Refs.~\cite{Tang:2002ww,Piasetzky:2006ai,JeffersonLabHallA:2007lly,Subedi:2008zz,LabHallA:2014wqo,CLAS:2020mom},
while the state of the residual nuclear system was not determined.
Hence, the reaction products may suffer incoherent rescattering processes in the residual nucleus, leading to the distortions of their momenta
and the formation of a highly excited nuclear residue, which may even be in an unbound state.

In the experiments performed at NIKHEF \cite{Onderwater:1997zz,Onderwater:1998zz,Starink:2000qhh} and MAMI \cite{ROSNER200099},
the reaction $^{16}\mbox{O}(e,e^\prime pp)^{14}\mbox{C}$ with production
of $^{14}\mbox{C}$ in the $0^+$ ground state and several excited states ($2^+$ at $E^*=7.01$ and 8.32 MeV, $0^+$ at $E^*=9.75$ MeV, and $1^+$ at $E^*=11.31$ MeV)
has been studied.
The main idea was to use specific final states of the outgoing nucleus as a filter for various reaction processes. For example, $^{14}$C in the $0^+$ states
associated with low recoil momentum selects the $^1S_0$ internal state of $pp$, while $^{14}$C in the $1^+$ state selects the $^3P$ states
of $pp$-pair. In Ref. \cite{Starink:2000qhh} the two independent theoretical analyses
within the Pavia \cite{Giusti:1997pa} and Gent \cite{Ryckebusch:1997gn} models have been performed concluding
that the ground state channel is well described by introducing central SRCs.
On the other hand, the $1^+$ channel needs to include the intermediate $\Delta$ excitation \cite{Ryckebusch:2003tu}, thus,
clearly demonstrating the importance of the selection of quantum state of residual nucleus for the observation of SRCs.

In Refs. \cite{Giusti:1999sv,Ryckebusch:1999xr}, the tensor correlations that influence the $^3S_1$ and $^3D_1$ $pn$ states
are found to be important for the $^{16}\mbox{O}(e,e^\prime pn)^{14}\mbox{N}$ exclusive channels with production of $^{14}\mbox{N}$ in the $1^+$ states.
However, since the energy resolution of the neutron detector \cite{Middleton:2007rr} turned out to be
not enough to resolve separate energy state of $^{14}\mbox{N}$, these important theoretical predictions still remain to be confirmed by new experiments.
In this sense, a more effective way to study $pn$ SRCs (which are more abundant as compared to the $pp$ ones) is to use reactions in the inverse kinematics
where the residual nucleus can also be detected, allowing the final state to be completely reconstructed without the need for neutron detection. 

The first (almost) fully exclusive measurements of the reactions $^{12}\mbox{C}(p,2pn_s)^{10}\mbox{B}$ and $^{12}\mbox{C}(p,2pp_s)^{10}\mbox{Be}$
in inverse kinematics with the collision of carbon nuclei with a momentum of 48 GeV/c with a proton target were recently performed
by the BM@N collaboration at JINR \cite{Patsyuk:2021fju}.
The main idea was to reduce distortions of the reaction kinematics due to initial- and final state interactions (ISI/FSI) by
detecting an unbroken nucleus in the final state. This allows to extract more clean information on the genuine SRC dynamics.   
An important feature of this experiment is the detection of a fast residual nucleus at a finite distance from the point of interaction
with the proton target.
Under these conditions, residual nuclei in short-lived particle-unstable states will not reach the detector.
Hence, only the ground state and low-lying excited states of the residual nucleus will contribute to the counting rate.
Therefore, it is extremely important that the appropriate theoretical formalism, in addition to the correct description of the SRC pair
and its motion relative to the residual nucleus, also takes into account individual transitions to low-lying excited states of the residual nucleus.

Theoretical modeling of SRCs is based on the separation of long-range mean-field interactions and (relatively) short-range residual two-body interactions,
which is confirmed by microscopic calculations \cite{Wiringa:2006ih,Wiringa:2013ala,Piarulli:2022ulk}.
It was shown \cite{Piarulli:2022ulk} that for nuclei with $A \leq 12$ the independent particle model (IPM) predicts the
number of the $NN$ pairs in different spin-isospin states $ST$ with an accuracy of about 10-15\% compared with calculation taking
into account correlations.
In this case, residual interactions primarily affect the internal wave function (WF) of the pair.
In Ref.~\cite{Wiringa:2006ih}, the binding energies of s- and p-shell nuclei were well described by a simple model in which the pion-exchange spin-isospin interaction
acts within $NN$-pairs of different $ST$ while counting of pairs in a given spin-isospin state was carried out within the IPM.

In the model of Refs.~\cite{Colle:2013nna,Ryckebusch:2018rct,Ryckebusch:2019oya}, SRCs were dynamically generated by correlation operators acting on the IPM WF.
The authors showed that the correlation operators (in the two-body cluster approximation) do not affect the distribution of center-of-mass (c.m.) momenta of correlated pairs,
while most of the SRCs is generated by the action of correlation operators on the $NN$ pairs in the lowest internal state of the relative radial quantum number $n=0$
and relative orbital momentum $l=0$.

Another SRC model is the generalized contact formalism (GCF), see Ref.~\cite{Cruz-Torres:2019fum} and refs. therein.  
In the GCF, the total number of SRCs is included through
nuclear contacts that can be extracted either from experiment \cite{Weiss:2014gua,Weiss:2016obx} or from microscopic calculations 
\cite{Weiss:2015mba,Cruz-Torres:2019fum}.
Internal WFs of the $NN$ SRCs in different spin-isospin states are zero-energy solutions of the
Schroedinger equation with either phenomenological or chiral two-body interactions.
The relative $NN-B$ (i.e. the c.m.) WF of SRCs is pure phenomenological.
Simulations using the GCF describe the BM@N data \cite{Patsyuk:2021fju} very well,
although this model does not include information about the internal WF of the residual nucleus and, therefore, implicitly
gives the total production rate for all possible internal states of the residual nuclear system.   

In this work, to resolve individual states of the residual nucleus, we use an approach based on the expansion
of the nuclear wave function in a series of states of the translationally-invariant harmonic oscillator shell model (TISM) \cite{NS}.
The basic properties of this approach were formulated in Ref.~\cite{Balashov:1964}
where the quasi-elastic knock-out of $d$-, $t$- and $\alpha$- clusters from $1p$-shell nuclei by fast
protons was studied.
Momentum distribution of clusters in the nucleus and spectrum of excitations of the residual nucleus after fragment ejection
were calculated in Ref.~\cite{Balashov:1964} using the shell model taking into account pair nucleon-nucleon correlations
which explicitly demonstrated the importance of the fractional parentage method.
Following Ref.~\cite{Balashov:1964} we assume that the number of SRC pairs
in the nucleus and the WF of the relative motion of $NN$-pair with respect to the c.m. of the residual nucleus are determined by the mean field,
while the  high-momentum part of the internal wave function of the $NN$-pair is governed by the short-range $NN$-interaction.
Accordingly, we use the TISM for the calculation of the spectroscopic amplitude
(see Eq.(\ref{S_A^X}) below) for separation of a two-nucleon cluster from the initial nucleus
and of the WF of relative motion of the cluster and the residual nucleus.
On the other hand, we rely on phenomenology to determine the intrinsic dynamic properties of SRC pairs.
Hence, for the $pn$ SRC pairs with $(S,T)=(1,0)$, the internal WF at high relative momenta ($\gtsim 0.4$ GeV/c) is identified with
the free deuteron WF, while for the $NN$ pairs with $(S,T)=(0,1)$ -- with the $^1S_0$ WF of free $pp$-scattering at zero energy.

In the TISM, the harmonic oscillator (HO) potential allows analytical calculation of the spectroscopic amplitudes and relative $NN-B$ WFs.
Residual $NN$-interactions beyond the HO potential are taken into account by mixing the TISM configurations in the intermediate coupling scheme
\cite{Balashov:1964,Boyarkina73}.
We should note here that within the TISM the type of the WF of the relative motion
of the cluster $X$ and the residual nucleus $B$  is determined by the antisymmetry property
of the internal WFs of the nuclei $A$ and $B$  and of the cluster $X$.
As a result, the number of the oscillator quanta $n$ corresponding to the relative $X-B$ motion is fixed as
\begin{equation}
  n = N_A - N_X - N_B~,     \label{nQuanta}
\end{equation} 
where $N_i$ is the number of oscillator quanta  corresponding to the internal motion in the nucleus/cluster $i$ ($i=A,B,X$).
This "oscillator rule" is widely used  in the theory of nucleon  clusters in light nuclei (see, for example, Refs.~\cite{Giusti:1999sv,Chant:1978zz}).
According to this rule, when a $NN$ pair is separated from the ground state of the $^{12}$C nucleus, the relative $NN-B$ motion has either
a $2S$ or $2D$ type for the transition to the residual nucleus in the configuration $s^4p^6$, or $0S$ type for transition to the $s^2p^8$ configuration.
On the  contrary, within the GCF \cite{Cruz-Torres:2019fum} the effects of antisymmetrization
are not involved  into consideration,  and for the WF of the  relative motion $NN-B$ an "averaged" Gaussian form is assumed.

Previously, the TISM spectroscopic approach was used to analyze the experimental data \cite{Albrecht:1979zy,Ero:1981zz}
on the quasi-elastic knock-out of fast deuterons from  light nuclei $^{6,7}$Li and $^{12}$C by protons at 670 MeV.
In these $(p,pd)$ processes, the large momentum transfer ($\sim 1.6$ GeV/c) from the proton beam to the knocked-out deuteron selects
the high-momentum component of the internal $pn$ WF associated with compact deuteron configurations. 
The TISM allows to describe the shapes of the measured spectra when absorptive initial- and final state interactions (ISI/FSI)
are taken into account \cite{Zhusupov:1986df,Zhusupov:1987sp}.

In Ref.~\cite{Albrecht:1978wy}, the quasi-elastic interaction with the dineutron $p \langle nn \rangle \to d+n$
has been experimentally studied in addition to the $p\langle pn \rangle \to d+p$.
In Ref.~\cite{Imambekov:1987}, the measured ratio of events $\langle nn \rangle / \langle pn \rangle$ has been reproduced reasonably well
assuming the dominance of the $\Delta$-isobar mechanism
of the $p \langle nN \rangle \to d+N$ process \cite{Imambekov:1986qt}.
The review of these works was done in \cite{Zhusupov:1987sp}.

The paper is organized as follows. In sec. \ref{model} we explain the underlying model starting from the impulse approximation (IA) amplitude.
The ISI/FSI are included in the eikonal approximation.   
Then, in sec. \ref{results} we compare our results with the BM@N data \cite{Patsyuk:2021fju}
and also give predictions for the absolute cross sections. Section \ref{summary} contains the summary of our results and possible further steps.

\section{The model}
\label{model}

\begin{figure}
  \begin{center}
    \includegraphics[scale = 0.50]{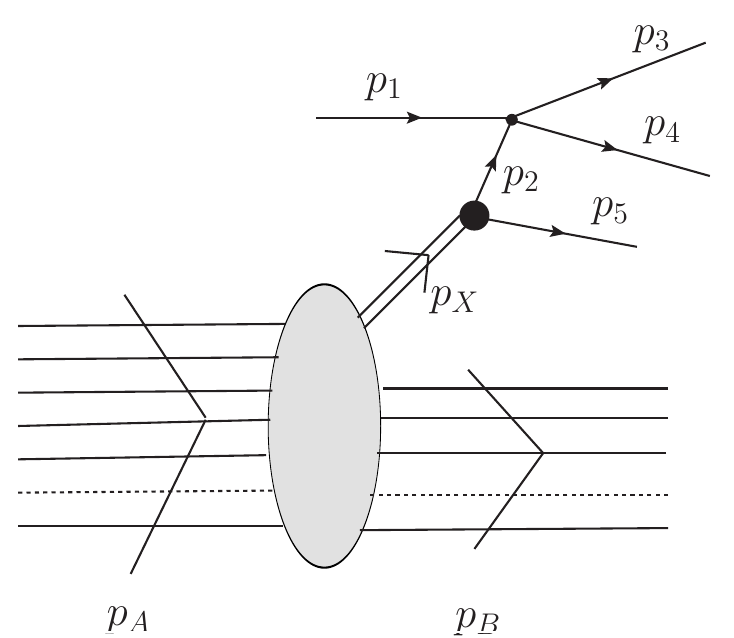}
  \end{center}
  \caption{\label{fig:diagr} The amplitude of the process $A(p,ppN)B$. 
    The lines are marked with four-momenta of the particles:
    the initial ($p_A$) and final ($p_B$) nuclei, beam proton ($p_1$), correlated $NN$ pair ($p_X$),
    struck proton ($p_2$), outgoing fast protons ($p_3$ and $p_4$), and outgoing slow nucleon ($p_5$).}
\end{figure}
The Feynman diagram of the IA amplitude is displayed in Fig.~\ref{fig:diagr}. For brevity, we use the notation ``$X$'' for the correlated $NN$-pair. 
The corresponding invariant matrix element is
then expressed as follows:
\begin{equation}
    M^{\rm IA} = M_{\rm hard}(p_3,p_4,p_1) 
              \frac{i\Gamma_{X \to pN}(p_{X},p_5)}{p_2^2-m^2+i\epsilon}
              \frac{i\Gamma_{A \to X B}(p_A,p_B)}{p_{X}^2-m_{X}^2+i\epsilon}~,  \label{M^IA}
\end{equation}
where $M_{\rm hard}(p_3,p_4,p_1)$ is the amplitude of hard elastic $pp$ scattering, $\Gamma_{A \to X B}(p_A,p_B)$ and
$\Gamma_{X \to pN}(p_{X},p_5)$ are the decay vertices.
Since the nucleus $B$ and nucleon $5$ are on the mass shell,
the decay vertices can be expressed in terms of the WFs of the relative motion in momentum space and particle energies
(see derivation in Ref. \cite{Larionov:2018lpk}):
\begin{eqnarray}
  && \frac{i\Gamma_{A \to X B}(p_A,p_B)}{p_{X}^2-m_{X}^2+i\epsilon}
  = S_A^X \left(\frac{2 E_B m_A}{p_{X}^0}\right)^{1/2} (2\pi)^{3/2} \psi_{n\Lambda}^{M_\Lambda}(-\bvec{p}_{X})~,  \label{Gamma_A}\\
  && \frac{i\Gamma_{X \to pN}(p_{X},p_5)}{p_2^2-m^2+i\epsilon}
  = \left(\frac{2 E_5 m_{X}}{p_2^0}\right)^{1/2} (2\pi)^{3/2} \sqrt{2} \psi_X(\bvec{p}_2)~, \label{Gamma_NN}
\end{eqnarray}
where all quantities on the r.h.s. are defined in the rest frame (r.f.) of decaying particle (i.e. of the nucleus $A$ in Eq.(\ref{Gamma_A}) and
of the $NN$-pair in Eq.(\ref{Gamma_NN})).
$\psi_{n\Lambda}^{M_\Lambda}(-\bvec{p}_{X})$ is the WF of the relative motion of the $NN$ pair and nucleus $B$, where
$n$ is the number of the oscillator quanta, $\Lambda$ is the orbital angular momentum, and $M_\Lambda$ is its $z$-component.
\footnote{Minus sign in the argument is related to the definition of the WF in the coordinate space, see Eq.(\ref{defFPC}) below.} 
$\psi_X(\bvec{p}_2)$ is the internal WF of the $NN$-pair. The factor $\sqrt{2}$ in the r.h.s. of Eq.(\ref{Gamma_NN})
comes from the antisymmetrized plane wave product WF of the decay nucleons.

The normalization conditions of the WFs are:
\begin{eqnarray}
  && \int d^3p |\psi_{n\Lambda}^{M_\Lambda}(\bvec{p})|^2 =1~, \label{norm1}\\
  && \int d^3p |\psi_X(\bvec{p})|^2 =1~,  \label{norm2}
\end{eqnarray} 
where in the last equation the sum over spin and isospin $z$-components of the decay nucleons is implicitly assumed.
In Eq.(\ref{Gamma_A}), $S_A^X$ is the spectroscopic amplitude of the transition from a given state of the nucleus $A$ to
a given state of the system $XB$, see Refs.~\cite{Smirnov:1977zz,Uzikov:2022epe}:
\begin{eqnarray}
  S_A^X &=&
  \left(\begin{array}{c}
    A \\
    2
  \end{array}\right)^{1/2} \langle \Psi_A | \Psi_B, n \Lambda, \Psi_X \rangle  \nonumber \\
        &=& \left(\begin{array}{c}
    A \\
    2
       \end{array}\right)^{1/2} \sum_{{\cal L} J_0 M_0}
   \left\{\begin{array}{lll}
              L_B      & S_B & J_B \\
              {\cal L} & S_X & J_0 \\
               L       & S   & J
          \end{array}
   \right\} \sqrt{(2L+1)(2S+1)(2J_B+1)(2J_0+1)}\, U(\Lambda L_X J_0 S_X;{\cal L} J_X)  \nonumber \\
  && \times \langle AN_A[f](\lambda\mu)\alpha LST|(A-2)N_B[f_B](\lambda_B\mu_B)\alpha_B  L_BS_BT_B; n\Lambda, 2 N_X[f_X](\lambda_X\mu_X)\alpha_X L_XS_XT_X\{{\cal L}\} \rangle \nonumber \\
  && \times  (J_B M_B J_0 M_0 | J M)\, (\Lambda M_\Lambda J_X M_X | J_0 M_0)\,
            (T_B M_{T_B} T_X M_{T_X} | T M_T)~.      \label{S_A^X}
\end{eqnarray}
Here, the internal WF of the initial nucleus $\Psi_A \equiv |A N_A[f](\lambda\mu)\alpha LST JMM_T \rangle$
is characterized by the following quantum numbers: 
$N_A$ -- the number of the oscillator quanta, $[f]$ -- the Young scheme,
$(\lambda\mu)$ -- the Elliott symbol,
$L$, $S$, $J$ -- the orbital, spin, and total angular momenta, respectively, $T$ -- isospin, $M$ and $M_T$ --  $z$-components of $J$ and $T$, respectively.
$\alpha$ denotes some possible additional quantum numbers needed for complete definition of the state.
Similar quantum numbers are also used to characterize the internal WFs of the final nucleus
$\Psi_B \equiv |(A-2)  N_B[f_B](\lambda_B\mu_B)\alpha_B L_BS_BT_B J_BM_BM_{T_B} \rangle$,
and $NN$-correlation $\Psi_X \equiv |2 N_X[f_X](\lambda_X\mu_X)\alpha_X L_XS_XT_X J_XM_XM_{T_X} \rangle$.
The numbers of oscillator quanta satisfy the oscillator rule, Eq.(\ref{nQuanta}).
In Eq.(\ref{S_A^X}), the standard notations for the $6j$- and $9j$-symbols are used, as defined in Ref.~\cite{VMKh},
while the factor $\langle \ldots \rangle$ is the fractional parentage coefficient (FPC) of the TISM.
The FPC enters the decomposition of the internal fully antisymmetric WF of the nucleus $A$ to the products
of fully antisymmetric internal WF of the nucleus $B$, WF of the relative motion of $B$
and $NN$-correlation $X$, and fully antisymmetric internal wave function of $X$:
\begin{eqnarray}
&&  |A N_A[f](\lambda\mu)\alpha LST M_L M_S M_T\rangle = \sum (L_B M_{L_B} {\cal L} M_{\cal L} | L M_L)\, (\Lambda M_\Lambda L_X M_{L_X} | {\cal L} M_{\cal L}) \nonumber \\
    && \times (S_B M_{S_B} S_X M_{S_X} | S  M_S)\, (T_B M_{T_B} T_X M_{T_X} | T M_T)  \nonumber \\
    && \times
  \langle A N_A[f](\lambda\mu)\alpha LST | (A-2) N_B[f_B](\lambda_B\mu_B)\alpha_B L_BS_BT_B;
   n\Lambda, 2 N_X[f_X](\lambda_X\mu_X)\alpha_X  L_XS_XT_X\{{\cal L}\} \rangle \nonumber \\
    && \times
   |(A-2)  N_B[f_B](\lambda_B\mu_B)\alpha_B L_BS_BT_B M_{L_B} M_{S_B} M_{T_B} \rangle \nonumber \\
   && \times \psi_{n\Lambda}^{M_\Lambda}(\bvec{R}_B - \bvec{R}_X)\, |2 N_X[f_X](\lambda_X\mu_X)\alpha_X L_XS_XT_X M_{L_X} M_{S_X} M_{T_X}\rangle~, \label{defFPC}
\end{eqnarray}
where the sum is taken over all quantum numbers apart those entering the WF of the nucleus $A$. In Eq.(\ref{defFPC}), the WF of the nucleus $B$ depends
on the fixed set of variables $x_1,x_2,\ldots,x_{A-2}$ while the WF of the correlation $X$ depends on the remaining variables
$x_{A-1}$ and $x_A$. Here, $x_i \equiv (\bvec{r}_i,\lambda_i,t_i)$ denote the position $\bvec{r}_i$, spin $\lambda_i$, and isospin $t_i$ variable of
the $i$-th nucleon. The c.m. coordinates of the nucleus $B$ and correlation $X$ are
\begin{equation}
  \bvec{R}_B = \frac{1}{A-2} \sum_{i=1}^{A-2} \bvec{r}_i  \label{R_B}
\end{equation}
and
\begin{equation}
  \bvec{R}_X = \frac{1}{2} (\bvec{r}_{A-1}+\bvec{r}_{A})~. \label{R_X}
\end{equation}
respectively. \footnote{Since, in Eq.(\ref{defFPC}), the WF of the nucleus $B$ is the internal one
it depends on $(A-3)$ Jacobi coordinates which can be defined as $\bvec{X}_1 = \bvec{r}_1 -\bvec{r}_2$, $\bvec{X}_2 = (\bvec{r}_1+\bvec{r}_2)/2 - \bvec{r}_3$,
...,$\bvec{X}_i = (\sum_{k=1}^i \bvec{r}_k)/i - \bvec{r}_{i+1}$,...,$\bvec{X}_{A-3} = (\sum_{k=1}^{A-3} \bvec{r}_k)/(A-3) - \bvec{r}_{A-2}$.
In a similar way, the internal WF of the correlation $X$ depends only the relative coordinate $\bvec{r}_{A-1}-\bvec{r}_{A}$.}

As discussed in sec.~\ref{intro}, only the ground state and low-lying excited states of the residual nucleus contribute to the reaction rate.
This excludes the excitation of the $\alpha$-core and selects the states of the residual nucleus with the minimum number of oscillator quanta.
The FPC of the TISM can be expressed via the FPC of the conventional shell model, and for the considered case
the following relationship is used \cite{Smirnov:1977zz}:
\begin{eqnarray}
&& \langle A N_A^{\rm min}[f](\lambda\mu) LST | (A-b) N_B^{\rm min}[f_B](\lambda_B\mu_B) L_BS_BT_B;
  n\Lambda, b N_X[f_X](\lambda_X\mu_X)  L_XS_XT_X\{{\cal L}\} \rangle     \nonumber \\
  && = (-1)^n \left(\frac{A}{A-b}\right)^{n/2}
  \left(\begin{array}{c}
    A-4 \\
    b
  \end{array}\right)^{1/2}
  \left(\begin{array}{c}
    A \\
    b
  \end{array}\right)^{-1/2}  \nonumber \\
  && \times \langle p^{A-4}[f](\lambda\mu) LST | p^{A-b-4}[f_B](\lambda_B\mu_B) L_BS_BT_B; p^b[f_X](\lambda_X\mu_X) {\cal L}S_XT_X \rangle \nonumber \\
  && \times  \langle p^b[f_X](\lambda_X\mu_X) {\cal L}S_XT_X | n\Lambda,  b N_X[f_X](\lambda_X\mu_X)  L_XS_XT_X \rangle~,  \label{FPCusual}
\end{eqnarray}
where $b$ is the number of nucleons in the cluster $X$. Eq.(\ref{FPCusual}) is derived by applying the Bethe-Rose-Elliott-Skyrme theorem
\cite{Bethe37,Elliott55} to the states of the nuclei
$A$ and $B$ which are supposed to contain the minimum numbers of the oscillator quanta
compatible with Pauli principle.
The last factor in the r.h.s. of Eq.(\ref{FPCusual}) is the cluster coefficient, i.e. the overlap integral of the shell model state of $b$ $p$-wave nucleons
with total angular momentum ${\cal L}$ and the product of the WF $\psi_{n\Lambda}(\bvec{R}_X)$ and the WF of the TISM $|b N_X[f_X](\lambda_X\mu_X)  L_XS_XT_X \rangle$.
It is assumed that the last two WFs are vector-coupled and have total angular momentum ${\cal L}$.

The expressions (\ref{S_A^X}),(\ref{FPCusual}) can be easily applied to calculate spectroscopic amplitudes using the two-particle FPCs of the conventional HO shell
model given in the tables of Refs.~\cite{Elliott53,NS}.
However, before to do this, we need to specify the nuclear states (we use the notation $^{(2T+1) (2S+1)}L$ below).

We will apply the shell model with intermediate coupling of Ref.~\cite{Boyarkina73} where the TISM basis states have been used to diagonalize the realistic nuclear Hamiltonian
that includes -- in addition to the HO one-body potential -- the Wigner (central), Majorana ($\propto \hat{P}_x$), Bartlett ($\propto \hat{P}_{\sigma}$)
and Heisenberg ($\propto  \hat{P}_x\hat{P}_{\sigma}$) two-body interactions and the spin-orbit one-body potential.  
Contributions of various TISM states to the energy eigenstates of light nuclei obtained in Ref.~\cite{Boyarkina73}
correspond to the qualitative assessments of Ref.~\cite{Wiringa:2006ih}.
The intermediate coupling model of Ref.~\cite{Boyarkina73} was also used by other authors
to calculate the quasi-elastic knock-out of $\alpha$-particles by protons \cite{Sakharuk:1997zz} and electrons \cite{Sakharuk:1999zz}.

The $^{12}$C ground state has quantum numbers $J=T=0$ and mainly consists of the $^{11}$S state with
maximum symmetry corresponding to the Young scheme $[44]$.
We neglect small contributions of the $[431]^{13}$P and other states.
In accordance with Ref.~\cite{Colle:2013nna} we will also require that the internal WF of the $NN$ SRC does not contain oscillator quanta,
i.e. $N_X=0$ and $L_X=0$, which corresponds to the choice of the most compact $NN$-configurations.
This choice determines the Young scheme $[f_X]=[2]$ and the Elliott symbol
$(\lambda_X\mu_X)=(20)$. Furthermore, for the minimum  oscillator quantum number
$N_B^{min}=6$  we have $n=2$ and  therefore $\Lambda=0,2$.
All this allows us to simplify Eq.(\ref{S_A^X}) to the following form:
\begin{eqnarray}
  S_A^X &=&
  \left(\begin{array}{c}
    A \\
    2
  \end{array}\right)^{1/2}  \langle AN_A[f](\lambda\mu) 000 |(A-2)N_B[f_B](\lambda_B\mu_B) L_BS_BT_B; n\Lambda, 2 N_X[f_X](\lambda_X\mu_X) 0 S_XT_X\{\Lambda\} \rangle
  \nonumber \\
  && \times \frac{\delta_{L_B\Lambda} \delta_{S_BS_X} \delta_{S_XJ_X} \delta_{T_BT_X} \delta_{M_{T_B},-M_{T_X}}}{\sqrt{(2L_B+1)(2S_B+1)(2T_B+1)}}
     (-1)^{J_B-M_B+T_B-M_{T_B}}\, (\Lambda M_\Lambda J_X M_X | J_B, -M_B)~. \label{S_A^X_simpl}
\end{eqnarray}
Thus, the allowed  values are $L_B=\Lambda=0,2$. The relevant FPCs are collected in Table~\ref{tab:FPC}
(for a more expanded set of FPCs, see Table~1 of Ref.~\cite{Uzikov:2022epe}),
and details of their calculation are given in Appendix~\ref{FPC}.
Note that the two-particle orbital FPCs allow coupling of the $[44]$ state of $^{12}$C to only $[42]$ or $[33]$ states of the residual  nucleus.
However, the $[33]$ state of the residual nucleus is coupled with the Young scheme $[11]$ of the $NN$ pair and, thus, can be discarded.
\begin{table}[htb]
  \caption{\label{tab:FPC}
    The FPCs of the TISM \\
    $\equiv \langle A N_A^{\rm min}[f](\lambda\mu) LST | (A-b) N_B^{\rm min}[f_B](\lambda_B\mu_B) L_BS_BT_B;
    n\Lambda, b N_X[f_X](\lambda_X\mu_X)  L_XS_XT_X\{{\cal L}\} \rangle$ for $A=12$, $N_A^{\rm min}=8$, $[f](\lambda\mu)=[44](04)$, $L=S=T=0$, $N_B^{\rm min}=6$,
    $[f_B](\lambda_B\mu_B)=[42](22)$, $b=2$, $N_X=0$, $[f_X](\lambda_X\mu_X)=[2](20)$, $L_X=0$, $S_X=S_B$, $T_X=T_B$.}
  \begin{center}
    \begin{tabular}{lll}
    \hline
    \hline
    $^{(2T_B+1)\,(2S_B+1)}L_B$ \hspace{1cm}   & $n~\Lambda$ \hspace{1cm} & FPC \\
    \hline
    $^{13}S$                 & 2 0        & $-\sqrt{8/275}$ \\
    $^{31}S$                 & 2 0        & $\sqrt{8/275}$ \\
    $^{13}D_{I}$             & 2 2         & $-\sqrt{3/550}$ \\
    $^{31}D_{I}$             & 2 2         & $\sqrt{3/550}$ \\
    $^{13}D_{II}$             & 2 2        & $-\sqrt{7/110}$ \\
    $^{31}D_{II}$             & 2 2         & $\sqrt{7/110}$ \\    
    \hline
    \hline
    \end{tabular}
  \end{center}
\end{table}

The outgoing nuclei $^{10}$B and $^{10}$Be can be in excited states. In the present exploratory study,
as explained in sec.~\ref{intro}, we will consider only a few low-lying excitations
consisting of different TISM states with minimum number of oscillator quanta.
The $^{10}$B and $^{10}$Be states included in our calculations, as well as the partial amplitudes of the contributing TISM states for $A=10$,
are collected in Tables~\ref{tab:10B} and \ref{tab:10Be} respectively.
The listed levels are confirmed by the compilation of experimental data in Ref.~\cite{Tilley:2004zz}.
All of the listed excited states are long-lived and can decay only due to the emission of $\gamma$.
(The only exception is the $^{10}$B $T=1$, $J=2$ state with $E^*=5.17$ MeV that has a photon decay branching ratio of $83\%$ with remaining $17\%$
in the $\alpha$-decay.)
\begin{table}[htb]
  \caption{\label{tab:10B} Experimental and theoretical (in parentheses) energy levels of $^{10}$B with the partial amplitudes of the TISM states with Young scheme $[42]$.
    Taken from Ref.~\cite{Boyarkina73}.}
    \begin{center}
    \begin{tabular}{llll}
    \hline
    \hline
    $E^*$, MeV   \hspace{1cm}  & $T~J$  \hspace{1cm}  &  TISM state \hspace{1cm} & $\alpha$ \\
    \hline
    0                          & 0 3                  &  $^{13}D_{I}$             & -0.418 \\
                               &                      &  $^{13}D_{II}$            & 0.679 \\
    0.717 (0.68)               & 0 1                  &  $^{13}S$                & -0.351 \\
                               &                      &  $^{13}D_{I}$             & 0.682 \\
                               &                      &  $^{13}D_{II}$            & 0.541 \\
    2.15 (2.08)                & 0 1                  &  $^{13}S$                & 0.885 \\
                               &                      &  $^{13}D_{I}$             & 0.307 \\
                               &                      &  $^{13}D_{II}$            & 0.224 \\
    3.58 (3.5)                 & 0 2                  &  $^{13}D_{I}$             & 0.401 \\  
                               &                      &  $^{13}D_{II}$            & 0.778 \\
    1.74 (1.51)                & 1 0                  &  $^{31}S$                & 0.772 \\
    5.17 (5.10)                & 1 2                  &  $^{31}D_{I}$             & 0.728 \\  
                               &                      &  $^{31}D_{II}$            & 0.209 \\
    \hline
    \hline
    \end{tabular}
  \end{center}
\end{table}
\begin{table}[htb]
  \caption{\label{tab:10Be} Same as in Table \ref{tab:10B} but for $^{10}$Be.}
    \begin{center}
    \begin{tabular}{llll}
    \hline
    \hline
    $E^*$, MeV   \hspace{1cm}  & $T~J$  \hspace{1cm}  &  TISM state \hspace{1cm} & $\alpha$ \\
    \hline
    0                          & 1 0                  &  $^{31}S$                & 0.772 \\
    3.368 (3.59)               & 1 2                  &  $^{31}D_{I}$             & 0.728 \\
                               &                      &  $^{31}D_{II}$            & 0.209 \\
    5.96  (5.96)               & 1 2                  &  $^{31}D_{I}$             & -0.226 \\
                               &                      &  $^{31}D_{II}$            & 0.892 \\
    \hline
    \hline
    \end{tabular}
  \end{center}
\end{table}
The total amplitude for the process  $^{12}\mbox{C}(p,2pN_s)B$ with the outgoing nucleus $B$ in a certain energy eigenstate is given by a coherent sum
\begin{equation}
  M_{\rm tot}^{\rm IA} = \sum_i M_i^{\rm IA} \alpha_i~,    \label{M_tot^IA}
\end{equation}
where $M_i^{\rm IA}$ is the amplitude (\ref{M^IA}) for the outgoing nucleus $B$ in the TISM eigenstate $i$, and $\alpha_i$ is the amplitude
of the state $i$ in the energy eigenstate, taken from Tables~\ref{tab:10B} and \ref{tab:10Be}.

ISI/FSI effects can be taken into account by replacing the incoming and outgoing plane waves with distorted waves.
In the eikonal approximation, the plane waves of the incoming proton ($i=1$) and outgoing nucleons ($i=3,4,5$) should be multiplied by
absorption factors (c.f. Refs.~\cite{Frankfurt:1994nn,Larionov:2017hcm})
\begin{equation}
  F_i(\bvec{r}) =
  \exp\left(-\frac{1}{2}\sigma_{NN}(p_i) T_i(\bvec{r})\right)~, \label{F_i}
\end{equation}
where $\sigma_{NN}(p_i)$ is the total $NN$ cross section depending on the momentum of the particle in the r.f. of the nucleus $B$.
\begin{equation}
  T_i(\bvec{r}) = \left\{  \begin{array}{lll}
                           \int\limits_{-\infty}^0 d\eta\, \rho(\bvec{r}+\hat{\bvec{p}}_i \eta)  &  {\rm for} & i=1 \\
                           \int\limits_0^{+\infty} d\eta\, \rho(\bvec{r}+\hat{\bvec{p}}_i \eta ) &  {\rm for} & i=3,4,5~.
                           \end{array}  \right.  \label{T_i}
\end{equation}
are the thickness functions with $\rho(\bvec{r})$ being the nucleon number density of the nucleus $B$ in the position $\bvec{r}$,
and $\hat{\bvec{p}} \equiv \bvec{p}/p$.
The absorption-corrected matrix element is obtained by substitution in Eq.(\ref{Gamma_A})
\begin{equation}
    (2\pi)^{3/2} \psi_{n\Lambda}^{M_\Lambda}(-\bvec{p}_{X})
    \to  \int d^3r \mbox{e}^{-i\bvec{p}_{X}\bvec{r}} \psi_{n\Lambda}^{M_\Lambda}(-\bvec{r})
    F_1(\bvec{r}) F_3(\bvec{r})  F_4(\bvec{r})  F_5(\bvec{r})~,~~~\bvec{r}=\bvec{R}_X-\bvec{R}_B~.       \label{FSI_corr}
\end{equation}
To summarize, we can write the following expression for the matrix element, which includes the summation over the magnetic quantum numbers of the intermediate states:
\begin{eqnarray}
  M_{\rm tot} &=& \sum_{\lambda_2} M_{\rm hard}(p_3,p_4,p_1)  \left[\left(\frac{2 E_5 m_{X}}{p_2^0}\right)^{1/2} (2\pi)^{3/2} \sqrt{2} \sum_{M_X} \psi_{X}(\bvec{p}_2)\right]_{{\rm r.f. of} X} \nonumber \\
  && \times \sum_i \alpha_i \sum_{M_\Lambda} S_{A,i}^{X} \left(\frac{2 E_B m_A}{p_{X}^0}\right)^{1/2} 
            \int d^3r\, \mbox{e}^{-i\bvec{p}_{X}\bvec{r}} \psi_{n_i\Lambda_i}^{M_\Lambda}(-\bvec{r}) F_{\rm abs}(\bvec{r})~,   \label{M_tot}
\end{eqnarray}
where $F_{\rm abs}(\bvec{r}) \equiv F_1(\bvec{r}) F_3(\bvec{r})  F_4(\bvec{r})  F_5(\bvec{r})$.
In Eq.(\ref{M_tot}), the factor in square brackets is evaluated in the r.f. of $NN$ correlation, and all other factors -- in the r.f. of the nucleus $A$.
For simplicity, we will further assume a spin-independent hard amplitude.
Note that in the nonrelativistic limit Eq.(\ref{M_tot}) corresponds to the simplified form of the eikonal approximation \cite{Janus:1974zz}
used in Ref.~\cite{Zhusupov:1986df} for the reaction $^{12}$C(p,pd)$^{10}$B
(except for four absorption factors $F_i({\bf r})$ in Eq.~(\ref{M_tot}) instead of three in Ref.~\cite{Zhusupov:1986df}).

Now we can calculate the modulus squared of the matrix element (\ref{M_tot}):
\begin{eqnarray}
  \overline{|M_{\rm tot}|^2} &\equiv&  \frac{1}{2} \sum_{\lambda_1,\lambda_3,\lambda_4,\lambda_5,M_B}  |M_{\rm tot}|^2 \nonumber \\
 &=& \overline{|M_{\rm hard}(p_3,p_4,p_1)|^2}   \left(\frac{2 E_5 m_{X}}{p_2^0}\right) (2\pi)^{3} 2 \overline{|\psi_{X}(\bvec{p}_2)|^2}  \nonumber \\
  && \times \sum_{i,j} \alpha_i \alpha_j \sum_{M_{\Lambda,i},M_{\Lambda,j}} \sum_{M_B,M_X} S_{A,i}^{X} S_{A,j}^{X}  \left(\frac{2 E_B m_A}{p_{X}^0}\right) \nonumber \\
    && \times \int d^3r \int d^3r^\prime \mbox{e}^{-i\bvec{p}_{X}(\bvec{r}-\bvec{r}^\prime)} \psi_{n_i\Lambda_i}^{M_{\Lambda,i}}(-\bvec{r}) \psi_{n_j\Lambda_j}^{M_{\Lambda,j}\,*}(-\bvec{r}^\prime)
    F_{\rm abs}(\bvec{r}) F_{\rm abs}(\bvec{r}^\prime)~,       \label{M_tot^2}
\end{eqnarray}
where we neglected the interference of amplitudes with different magnetic quantum numbers
$M_X$ of the $NN$-pair and eliminated the spin correlations between $\psi_{X}(\bvec{p}_2)$ and $M_{\rm hard}(p_3,p_4,p_1)$
by successive replacements
\begin{eqnarray}
  \frac{1}{2} \sum_{\lambda_1,\lambda_3,\lambda_4} |M_{\rm hard}(p_3,p_4,p_1)|^2 & \to & \overline{|M_{\rm hard}(p_3,p_4,p_1)|^2}
  \equiv \frac{1}{4} \sum_{\lambda_1,\lambda_2,\lambda_3,\lambda_4} |M_{\rm hard}(p_3,p_4,p_1)|^2~,              \label{overlineMhard2def} \\
  \sum_{\lambda_2,\lambda_5} |\psi_{X}(\bvec{p}_2)|^2 & \to & \overline{|\psi_{X}(\bvec{p}_2)|^2}
  \equiv \frac{1}{2J_X+1}\sum_{M_X,\lambda_2,\lambda_5} |\psi_{X}(\bvec{p}_2)|^2~.                          \label{overlineWF2def}
\end{eqnarray}
By using Eq.(\ref{S_A^X_simpl}) and the property of the Clebsch-Gordan coefficients 
\begin{equation}
  \sum_{M_B,M_X} (\Lambda_i M_{\Lambda,i} J_X M_X | J_B, -M_B) (\Lambda_j M_{\Lambda,j} J_X M_X | J_B, -M_B)
  = \frac{2J_B+1}{2\Lambda_i+1} \delta_{\Lambda_i,\Lambda_j} \delta_{M_{\Lambda,i},M_{\Lambda,j}}     \label{CGprop}
\end{equation}
we can simplify Eq. (\ref{M_tot^2}) as follows
\begin{eqnarray}
  \overline{|M_{\rm tot}|^2} &=& (2J_B+1) \overline{|M_{\rm hard}(p_3,p_4,p_1)|^2} \left(\frac{2 E_5 m_{X}}{p_2^0}\right) (2\pi)^{3} 2 \overline{|\psi_{X}(\bvec{p}_2)|^2} \nonumber \\
  && \times \sum_{i,j} \alpha_i \alpha_j  \delta_{\Lambda_i,\Lambda_j} S_{A,i}^{X0} S_{A,j}^{X0}  \left(\frac{2 E_B m_A}{p_{X}^0}\right) \nonumber \\
  && \times \int d^3r \int d^3r^\prime \mbox{e}^{-i\bvec{p}_{X}(\bvec{r}-\bvec{r}^\prime)} \frac{1}{2\Lambda_i+1} \sum_{M_\Lambda} \psi_{n_i\Lambda_i}^{M_{\Lambda}}(-\bvec{r})
                                                                                                                   \psi_{n_i\Lambda_i}^{M_{\Lambda}\, *}(-\bvec{r}^\prime)
  F_{\rm abs}(\bvec{r}) F_{\rm abs}(\bvec{r}^\prime)~,        \label{M_tot^2_simp}
\end{eqnarray}
where we introduced the reduced spectroscopic amplitude 
\begin{equation}
     S_{A,i}^{X0} \equiv 
\left(\begin{array}{c}
    A \\
    2
  \end{array}\right)^{1/2} 
\frac{FPC_i}{\sqrt{(2L_B+1)(2S_B+1)(2T_B+1)}} \label{S_A^X0}
\end{equation}
and also used the fact that the value of $n_i$ is fixed for the selected set of FPCs.

It is convenient to perform the double space integration in Eq.(\ref{M_tot^2_simp}) in the variables 
$\bvec{R} \equiv (\bvec{r}+\bvec{r}^\prime)/2$, $\bvec{\xi} \equiv \bvec{r}-\bvec{r}^\prime$.
In the spirit of the shadowed multiple scattering in Glauber theory \cite{Glauber:1970jm},
the double space integral in Eq.(\ref{M_tot^2_simp}) can be then approximately expressed as follows:
\begin{equation}
  \int d^3R \int d^3\xi \mbox{e}^{-i\bvec{p}_{X}\bvec{\xi}} \frac{1}{2\Lambda_i+1} \sum_{M_\Lambda} \psi_{n_i\Lambda_i}^{M_{\Lambda}}(-\bvec{R}-\bvec{\xi}/2) \psi_{n_i\Lambda_i}^{M_{\Lambda}\, *}(-\bvec{R}+\bvec{\xi}/2)
  F_{\rm abs}^2(\bvec{R})~, \label{DspaceInt}
\end{equation}
where we replaced $\bvec{r},\bvec{r}^\prime \to \bvec{R}$ in the arguments of the absorption factors $F_{\rm abs}$.
This approximation is valid because $F_{\rm abs}(\bvec{r})$ varies with $\bvec{r}$ on a relatively large length scale $2/\rho_0 \sigma_{NN} \simeq 3$ fm,
where $\rho_0 \simeq 0.16$ fm$^{-3}$ is the nucleon saturation density, and $\sigma_{NN} \simeq 40$ mb is the $NN$ cross section.
\footnote{Due to absorption factors, the integral in Eq.(\ref{M_tot^2_simp}) is dominated by a peripheral nuclear region where
the nucleon density is smaller than  $\rho_0$. Thus, the actual length scale may even be larger.} 
In contrast, the WF of the relative $X-B$ motion varies on a shorter length scale of the HO parameter
$r_0 \simeq 1.6$ fm (c.f. Ref.~\cite{Alkhazov:1972oie}) and, moreover, may contain nodes and/or non-monotonic behavior as a function of $R$.
Thus, we keep the exact arguments in the WFs of Eq.(\ref{DspaceInt}), allowing it to be rewritten as follows:
\begin{equation}
  \int d^3R f_{n_i\Lambda_i}(-\bvec{R},-\bvec{p}_{X})  F_{\rm abs}^2(\bvec{R})~, \label{DspaceInt1}
\end{equation}
where
\begin{equation}
  f_{n_i\Lambda_i}(-\bvec{R},-\bvec{p}_{X}) \equiv  \int d^3\xi \mbox{e}^{-i\bvec{p}_{X}\bvec{\xi}} \frac{1}{2\Lambda_i+1} \sum_{M_\Lambda} \psi_{n_i\Lambda_i}^{M_{\Lambda}}(-\bvec{R}-\bvec{\xi}/2)
  \psi_{n_i\Lambda_i}^{M_{\Lambda}\, *}(-\bvec{R}+\bvec{\xi}/2)    \label{WignerFunction}
\end{equation}
is a Wigner function which has a meaning of the probability density in the phase space $(\bvec{R},\bvec{p}_{X})$ where $\bvec{R}=\bvec{R}_X-\bvec{R}_B$
is the relative position and $\bvec{p}_{X}=-\bvec{p}_{B}$ is the canonically conjugated momentum. The Wigner function satisfies the relations:
\begin{eqnarray}
  && \int \frac{d^3R}{(2\pi)^3} f_{n_i\Lambda_i}(-\bvec{R},-\bvec{p}_{X}) = \overline{|\psi_{n_i\Lambda_i}(-\bvec{p}_{X})|^2}~,    \label{rel1} \nonumber \\
  && \int \frac{d^3p_X}{(2\pi)^3} f_{n_i\Lambda_i}(-\bvec{R},-\bvec{p}_{X}) = \overline{|\psi_{n_i\Lambda_i}(-\bvec{R})|^2}~,      \label{rel2} 
\end{eqnarray}
where overline means averaging over $M_{\Lambda}$.

As a result, we arrive at the following formula for the modulus squared of the matrix element:
\begin{eqnarray}
  \overline{|M_{\rm tot}|^2} &=& (2J_B+1) \overline{|M_{\rm hard}(p_3,p_4,p_1)|^2} \left(\frac{2 E_5 m_{X}}{p_2^0}\right) (2\pi)^{3} 2 \overline{|\psi_{X}(\bvec{p}_2)|^2} \nonumber \\
  && \times \sum_{i,j} \alpha_i \alpha_j  \delta_{\Lambda_i,\Lambda_j} S_{A,i}^{X0} S_{A,j}^{X0}  \left(\frac{2 E_B m_A}{p_{X}^0}\right)
     \int d^3R\,  f_{n_i\Lambda_i}(-\bvec{R},-\bvec{p}_{X}) F_{\rm abs}^2(\bvec{R})~, \label{M_tot^2_fin}
\end{eqnarray}
where the square of the absorption factor is
\begin{equation}
  F_{\rm abs}^2(\bvec{R}) = \exp[-\sigma_{NN}(p_1) T_1(\bvec{R}) - \sigma_{NN}(p_3) T_3(\bvec{R}) - \sigma_{NN}(p_4) T_4(\bvec{R}) - \sigma_{NN}(p_5) T_5(\bvec{R})]~. \label{F_abs^2}
\end{equation}
Thus, the way the absorption enters into Eq.(\ref{M_tot^2_fin}) has a simple and clear meaning:
the partial reaction rate in the IA for the c.m. of the
$NN$ correlation located in the space element $d^3R$ relative to the center of the residual nucleus $B$
is multiplied by the probability that the incoming proton will reach the point $\bvec{R}$ without being absorbed
(i.e. does not participate in any elastic or inelastic scattering processes) and
the outgoing nucleons will reach free space from the point $\bvec{R}$ without being absorbed.
We have checked that for the cross sections integrated over momentum of the
residual nucleus the approximate formula of Eq.(\ref{M_tot^2_fin}) agrees with Eq.(\ref{M_tot^2}) with accuracy of $\sim 20\%$.
This is acceptable for our purposes in this work
given the fact that the cross section is reduced by absorption effects by an order of magnitude.

The formula (\ref{M_tot^2_fin}) can be simply modified to estimate the contribution of charge exchange (CEX) processes.
Since the probability of collision with $pp$-correlation is small, we will take into account only two dominant contributions:
(i) when the incoming proton (1) interacts with the proton of $pn$-correlation, and the slow recoil neutron (5) then experiences CEX on a
proton from the nucleus $B$,
and (ii) when the incoming proton interacts with the neutron of $pn$-correlation, and the fast knocked-out neutron (4) then experiences CEX.
\footnote{In the second case, we assume that $\overline{|M_{\rm hard}|^2}$ is isospin-independent and both the 3-rd and 4-th nucleons can be neutrons
with probability 1/2, but the probability of CEX is the same for each of them.}
In the semi-classical approximation, the corresponding CEX probabilities are determined by the following expressions:
\begin{eqnarray}
    && P_{n_5 \to p_5} = F_{\rm abs}^2(\bvec{R}) \sigma_{\rm CEX}(p_5) T_5(\bvec{R})/2~, \\
    && P_{n_4 \to p_4} = F_{\rm abs}^2(\bvec{R}) \sigma_{\rm CEX}(p_4) T_4(\bvec{R})/2~.
\end{eqnarray}
where $1/2$ factors are included to obtain proton thickness functions. The total CEX probability is:
\begin{equation}
    P_{n_5 \to p_5} + P_{n_4 \to p_4} =  F_{\rm abs}^2(\bvec{R}) [\sigma_{\rm CEX}(p_5) T_5(\bvec{R}) + \sigma_{\rm CEX}(p_4) T_4(\bvec{R})]/2~.  \label{P_CEX}
\end{equation}
Similar expressions for CEX probabilities are used in GCF calculations, see Ref.~\cite{CLAS:2018xvc}.
To obtain the total reaction rate with outgoing slow proton we sum-up the reaction rate of Eq.(\ref{M_tot^2_fin}) for the $pp$-correlation and the
reaction rate of Eq.(\ref{M_tot^2_fin}) for the $pn$-correlation with CEX (i.e. with $F_{\rm abs}^2(\bvec{R})$ replaced by $P_{n_5 \to p_5} + P_{n_4 \to p_4}$). 
The back reactions $p \to n$ and the double CEX processes are neglected.

\subsection{Wave functions}
\label{WFS}

So far we have assumed that all WFs of our model are the TISM WFs.
However, phenomenology favors the deuteron-like WFs of isoscalar SRCs.
Thus, for isoscalar $pn$-pairs we use the deuteron WF of the CD-Bonn model, Ref.~\cite{Machleidt:2000ge}, which gives:
\begin{equation}
  \overline{|\psi_{X,T=0}(\bvec{p}_2)|^2} =  \frac{1}{2}\, \frac{u^2(p_2)+w^2(p_2)}{4\pi}~.       \label{DWF2}
\end{equation}
Here, $u(p_2)$ and $w(p_2)$ are, respectively, the $S$- and $D$-wave components satisfying normalization condition
\begin{equation}
   \int dp\, p^2 [u^2(p)+w^2(p)] = 1~.       \label{normDWF}
\end{equation}
The factor $1/2$ in Eq.(\ref{DWF2}) comes from the square of the isospin WF. 
For the free $^1S_0$ $pp$ ($T_z=1$) and $pn$ ($T_z=0$) pairs with $T=1$, the bound state is absent
but there is a virtual level (the pole of $S$-matrix) in the non-physical region of the
relative energy $E$ at  $E\approx-0.45$MeV.
Thus, for isovector $NN$-pairs we rely on the following formula:
\begin{equation}
  \overline{|\psi_{X,T=1}(\bvec{p}_2)|^2} = \frac{1}{2}(1+T_z) |\psi_s(p_2)|^2~.    \label{zeroEnSol2}
\end{equation}
Here, $\psi_s(p)$ is the zero-energy solution of $NN(^1S_0)$ scattering problem, Ref.~\cite{Uzikov:2022epe},
\begin{equation}
   \psi_s(p) = \kappa \frac{f(p,0;0)}{\alpha^2+p^2}~,          \label{zeroEnSol}
\end{equation}
where $\alpha=0.104$ fm$^{-1}$ corresponds to a virtual level ``binding energy'' $-E =\alpha^2/m=0.45$ MeV, Ref.~\cite{Faeldt:1996na}.
$f(p,k;k)$ is the half-off-shell $NN$ scattering amplitude in the $^1S_0$ channel parameterized
in Ref.~\cite{Lensky:2005hb}.
The factor $\kappa$ is chosen from the normalization condition:
\begin{equation}
  4\pi \int dp\, p^2 |\psi_s(p)|^2 = 1~.     \label{normZeroEnSol}
\end{equation}

The squared WFs of the relative motion $NN-B$ for the oscillator quantum number $n=2$ are given by
standard expressions from TISM:
\begin{eqnarray}
  |\psi_{20}(R)|^2             &=& \frac{3}{2 R_0^3 \pi^{3/2}}
                                 \left[1 - \frac{2}{3}\left(\frac{R}{R_0}\right)^2\right]^2 \exp[-(R/R_0)^2]~,    \label{psi_20^2} \\
  \overline{|\psi_{22}(R)|^2} &=& \frac{4}{15 R_0^3 \pi^{3/2}} \left(\frac{R}{R_0}\right)^4 \exp[-(R/R_0)^2]~,       \label{psi_22^2}
\end{eqnarray}
where $R_0=r_0\sqrt{A/2(A-2)}$ is the HO parameter of the $NN-B$ relative motion, and $r_0=1.736$ fm is the conventional HO model parameter fit to describe
the momentum distributions of the p-shell and s-shell nucleons in the $^{12}\mbox{C}(e,e^\prime p)^{11}\mbox{B}$ reaction \cite{Uzikov:2021wrh}.
\footnote{The conventional shell model fit to the differential cross section of elastic $p^{12}$C scattering at 1 GeV
gives $r_0=1.581$ fm \cite{Alkhazov:1972oie} which does not lead to significant changes in our numerical results.} 
We have also performed calculations using the phenomenological TISM
WF of the lowest HO state $n=0,\Lambda=0$
\begin{equation}
   |\psi_{00}(R)|^2 = \frac{1}{R_0^3 \pi^{3/2}} \exp[-(R/R_0)^2]~,   \label{psi_00^2}    
\end{equation}
where $R_0=\sqrt{\alpha_{cm}}=1$ fm \cite{CiofidegliAtti:1995qe}. This corresponds to the standard deviation
of the relative $NN-B$ momentum distribution $\sigma_{cm} =1/\sqrt{2\alpha_{cm}} = 139.5$ MeV/c,
which is consistent with $\sigma_{cm}=(156 \pm 27)$ MeV/c obtained from analysis of BM@N data \cite{Patsyuk:2021fju}.

According to the oscillator rule, Eq.(\ref{nQuanta}), WFs with $n=2$
correspond to transitions into $s^4p^{6}$  configurations, and $n=0$ -- into $s^2p^{8}$ configurations of the residual nucleus.
Since the latter configurations, most likely, are not included
into the BM@N data in question, taking $n=0$ is not allowed by the oscillator rule.
However, we  consider here this option too for comparison  with other SRC models
where this option  is often used.

The normalization of the $NN-B$ WFs is such that
\begin{equation}
  4\pi \int dR\, R^2\,  \overline{|\psi_{n\Lambda}(R)|^2} =1~.       \label{normNNBWFs}
\end{equation}
The transition to the WFs in momentum space  is simply reached by replacing $R \to p_X$ and $R_0 \to 1/R_0$ in Eqs.(\ref{psi_20^2}),(\ref{psi_22^2}),(\ref{psi_00^2}).

For the calculations with absorption, we need to specify the Wigner functions, Eq.(\ref{WignerFunction}). After somewhat lengthy but straightforward
calculations we arrive at the following formulas:
\begin{eqnarray}
  f_{20}(-\bvec{R},-\bvec{p}_X) &=& 8\,\mbox{e}^{-(R^2+p_X^2 R_0^4)/R_0^2} \left[ \frac{2}{3R_0^4}(R^4+p_X^4 R_0^8) 
    - \frac{4}{3} R^2 p_X^2 + 1 - \frac{4}{3R_0^2}(R^2+p_X^2 R_0^4) \right.  \nonumber \\
    && \left. + \frac{8}{3} (\bvec{R} \bvec{p}_X)^2 \right]~, \label{f_20} \\
  f_{22}(-\bvec{R},-\bvec{p}_X) &=& 8\,\mbox{e}^{-(R^2+p_X^2 R_0^4)/R_0^2} \left[ \frac{4}{15R_0^4}(R^4+p_X^4 R_0^8)
    +  \frac{16}{15} R^2 p_X^2 + 1 - \frac{4}{3R_0^2}(R^2+p_X^2 R_0^4) \right.  \nonumber \\
    && \left. - \frac{8}{15} (\bvec{R} \bvec{p}_X)^2 \right]~, \label{f_22} \\
  f_{00}(-\bvec{R},-\bvec{p}_X) &=& 8\,\mbox{e}^{-(R^2+p_X^2 R_0^4)/R_0^2}~.   \label{f_00}
\end{eqnarray}

\subsection{Elementary cross sections}
\label{elem}

Experimental data on $pp$ elastic large-angle differential cross section $d\sigma/d\Omega_{\rm c.m.}$ at $p_{\rm lab}=4$ GeV/c are reported in Ref.~\cite{Kammerud:1971ac}
as a function of the c.m. polar scattering angle $\Theta_{\rm c.m.}$. To get the square of the hard scattering amplitude, we used a simple relation
\begin{equation}
     \overline{|M_{\rm hard}(p_3,p_4,p_1)|^2} = 64 \pi^2 s\, d\sigma/d\Omega_{\rm c.m.}~,     \label{M_hard^2} 
\end{equation}
where $\Theta_{\rm c.m.}=\arccos[1+\max(t,u)/2(s/4-m^2)]$, $s=(p_3+p_4)^2$, $t=(p_1-p_3)^2$, $u=(p_1-p_4)^2$.

The experimental total $pp$ and $pn$ cross sections at the beam momentum $p < 5$ GeV/c are well described by the  parameterization of  Ref.~\cite{Cugnon:1996kh}
which we use in calculation of the absorption factor (\ref{F_abs^2}) with appropriate weighting according to the proton and neutron numbers so that 
the $pN$ and $nN$ cross sections are
\begin{eqnarray}
   \sigma_{pN} &=& [\sigma_{pp} Z + \sigma_{pn} (A-Z)]/A~,  \label{sigma_pN}\\
   \sigma_{nN} &=& [\sigma_{pn} Z + \sigma_{pp} (A-Z)]/A~,  \label{sigma_nN}
\end{eqnarray}
where $Z$ and $A$ are, respectively, the charge and mass numbers of the residual nucleus $B$. 

The CEX cross section $np \to pn$ is defined as the integrated elastic $np$ differential cross section
at large $\Theta_{\rm c.m.}$, typically at $\Theta_{\rm c.m.} > 90\degree$. Experimental data of Ref.~\cite{Jain:1984ps}
at $E=800$ MeV give $\sigma_{\rm CEX}(800)=4.25$ mb for $\Theta_{\rm c.m.}=135\degree - 180\degree$. 
The CEX cross section at other energies can be determined from the scaling relation
valid for $p_{\rm lab} < 100$ GeV/c, established in Ref.~\cite{Gibbs:1994um}:
\begin{equation}
  \sigma_{\rm CEX}(E)=\sigma_{\rm CEX}(800) \frac{s(800)}{s(E)}~,     \label{sigma_CEX}
\end{equation}
where $s(E)=2m(E+2m)$.

\subsection{Observables}
\label{observ}

The full differential cross section of the reaction $A(p,2pN_s)B$ is given by the standard formula (see Fig.~\ref{fig:diagr}
for particle notation):
\begin{equation}
   d\sigma_{1A \to 345B} = \frac{(2\pi)^4 \overline{|M_{\rm tot}|^2}}{4I} d\Phi_4~,     \label{dsigma_1A345B}
\end{equation}
where $I=[(p_1p_A)^2-m^2m_A^2]^{1/2}$ is the flux factor and
\begin{equation}
  d\Phi_4 = \delta^{(4)}(p_1+p_A-p_3-p_4-p_5-p_B) \frac{d^3p_3}{(2\pi)^32E_3} \frac{d^3p_4}{(2\pi)^32E_4}  \frac{d^3p_5}{(2\pi)^32E_5} \frac{d^3p_B}{(2\pi)^32E_B}
                                             \label{dPhi_4}
\end{equation}
is the four-body invariant phase space. In order to perform comparison with the BM@N data \cite{Patsyuk:2021fju}, we have to integrate Eq.(\ref{dsigma_1A345B})
over full four-body phase space applying experimental cuts. It is convenient to perform this in the following way.
First, we separate the two-body phase space $d\Phi_2$ of the outgoing fast protons:
\begin{equation}
   d\Phi_2 = \delta^{(4)}({\cal P}-p_3-p_4) \frac{d^3p_3}{(2\pi)^32E_3} \frac{d^3p_4}{(2\pi)^32E_4}~,   \label{dPhi_2}
\end{equation}
where ${\cal P}=p_1+p_A-p_5-p_B$. Integrating Eq.(\ref{dPhi_2}) over $d^3p_4$ and $dp_3$ gives the following result:
\begin{equation}
   d\Phi_2 = \frac{p_3 d\Omega_3}{(2\pi)^6 4 |E_3+E_4-E_3{\cal P}\chi/p_3|}~,       \label{dPhi_2_int}
\end{equation}
where $d\Omega_3$ is the solid angle of the 3-rd proton and $\chi \equiv \bvec{{\cal P}} \bvec{p}_3/{\cal P}p_3$.
All quantities in Eq.(\ref{dPhi_2_int}) are defined in the laboratory frame, i.e. in the r.f. of the 1st proton.
The momentum of the 3-rd proton is found by solving the equation $E_3+E_4={\cal P}^0$, which gives two solutions:
\begin{equation}
  p_3 = \frac{M^2{\cal P}\chi \pm {\cal P}^0 \sqrt{M^4 - 4m^2[({\cal P}^0)^2 -  \bvec{{\cal P}}^2\chi^2]}}%
    {2[({\cal P}^0)^2 - \bvec{{\cal P}}^2\chi^2]}~,          \label{p_3}
\end{equation}
with $M = \sqrt{{\cal P}^2}$ being the invariant mass of the 3-rd and 4-th protons. (In case, if both values of $p_3$
pass the BM@N acceptance, the r.h.s of Eq.(\ref{dPhi_2_int}) is summed over these two values.)

The integrations of Eq.(\ref{dPhi_4}) over three-momenta of the 5-th nucleon and residual nucleus should be performed in the r.f. of
$^{12}$C since the internal WF of the $NN$-correlation and the WF of relative $NN-B$ motion favor small momenta in that frame.
Replacing integration over $d^3p_5$ by integration over $d^3p_2$ (which is more convenient since the BM@N acceptance
restricts $|\bvec{p}_2|$) we come to the following formula for the integrated cross section:
\begin{equation}
  \sigma_{1A \to 345B} = \frac{1}{64(2\pi)^8p_{\rm beam}m} \int d\Omega_3 \int d^3p_2 \int \frac{d^3p_B}{E_5E_B}
  \frac{\overline{|M_{\rm tot}|^2} p_3}{|E_3+E_4-E_3{\cal P}\chi/p_3|}~,     \label{sigma_1A345B}
\end{equation}
where the energies $E_5$ and $E_B$ are defined in the r.f. of $^{12}$C, and $p_{\rm beam}$ is the momentum of $^{12}$C
in the lab. frame.
The single differential cross sections $d\sigma/dx$ where $x$ is any kinematic observable
are obtained by multiplying the integrand of Eq.(\ref{sigma_1A345B}) by $\delta(x-x(\Omega_3,\bvec{p}_2,\bvec{p}_B))$. 

Let us now summarize the BM@N acceptance cuts which are included when taking the integrals in Eq.(\ref{sigma_1A345B}).
\begin{itemize}
    \item Velocities of fast protons in the lab. frame: $0.8 < \beta_{3,4} < 0.96$.
    \item Polar angles of fast protons in the lab. frame: $24\degree < \Theta_{3,4} < 37\degree$.
    \item Azimuthal angles of fast protons in the lab. frame:  $-14\degree < \phi_3 < 14\degree$, $-180\degree < \phi_4 < -166\degree$ and $166\degree < \phi_4 < 180\degree$.
    \item In-plane opening angle: $\Theta_3+\Theta_4 > 63\degree$.
    \item Missing momentum in the r.f. of $^{12}$C: $0.350~\mbox{GeV/c} < p_2 < 1.2~\mbox{GeV/c}$.
    \item Missing energy $E_{\rm miss} \equiv m-p_2^0$ in the r.f. of $^{12}$C: $-0.110~\mbox{GeV} < E_{\rm miss} < 0.240~\mbox{GeV}$.
\end{itemize}

The calculation of the integrated cross section, Eq.(\ref{sigma_1A345B}), includes the eight-dimensional integral over phase space.
In the case, if absorption is included in the modulus squared of the matrix element, the integral becomes eleven-dimensional
\footnote{The thickness functions, Eq.(\ref{T_i}), have been calculated analytically by using the HO density profile of the residual nucleus
$
   \rho(r) = \frac{4}{r_0^3 \pi^{3/2}} \left[1 + \frac{A_B-4}{6} \left(\frac{r}{r_0}\right)^2\right] \mbox{e}^{-r^2/r_0^2}
$.} 
which makes the direct numerical calculation impossible. In order to overcome this problem, we have tabulated the absorption integral, Eq.(\ref{DspaceInt1}),
as a function of the momentum of slow nucleon, $\bvec{p}_5$. The $\bvec{p}_{X}$-dependence has been factorized out by using explicit forms, Eqs.(\ref{f_20})-(\ref{f_00}),
of Wigner functions. The momenta of fast protons have been fixed by the condition of free $pp$ scattering
at $\Theta_{c.m.}=90\degree$ which approximately corresponds to the middle of experimental acceptance region.
By selecting different kinematics within detector acceptance we have checked that this approximation works very well.

\section{Results}
\label{results}

In this section we present the results of our calculations of various single differential cross sections
of the reaction channel $^{12}\mbox{C}(p,2pn_s) ^{10}\mbox{B}$ compared to BM@N data from Ref.~\cite{Patsyuk:2021fju}.
Data points are scaled by conveniently chosen factors to facilitate comparison of measured distribution shapes with calculated ones.

\begin{figure}
  \begin{center}
  \includegraphics[scale = 0.60]{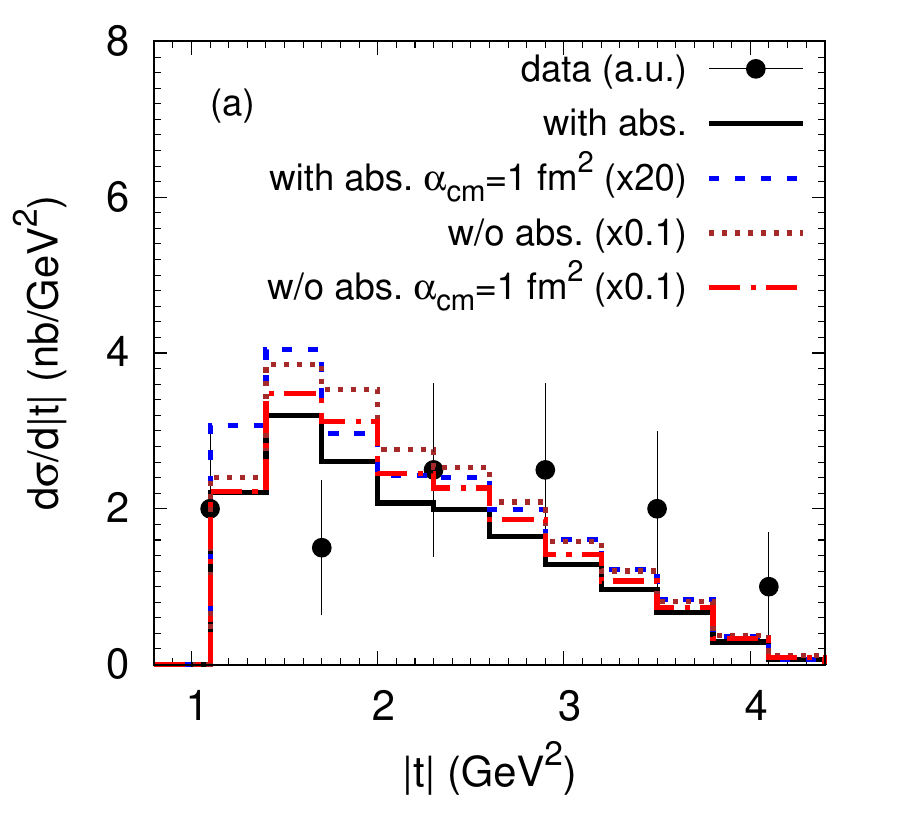}
  \includegraphics[scale = 0.60]{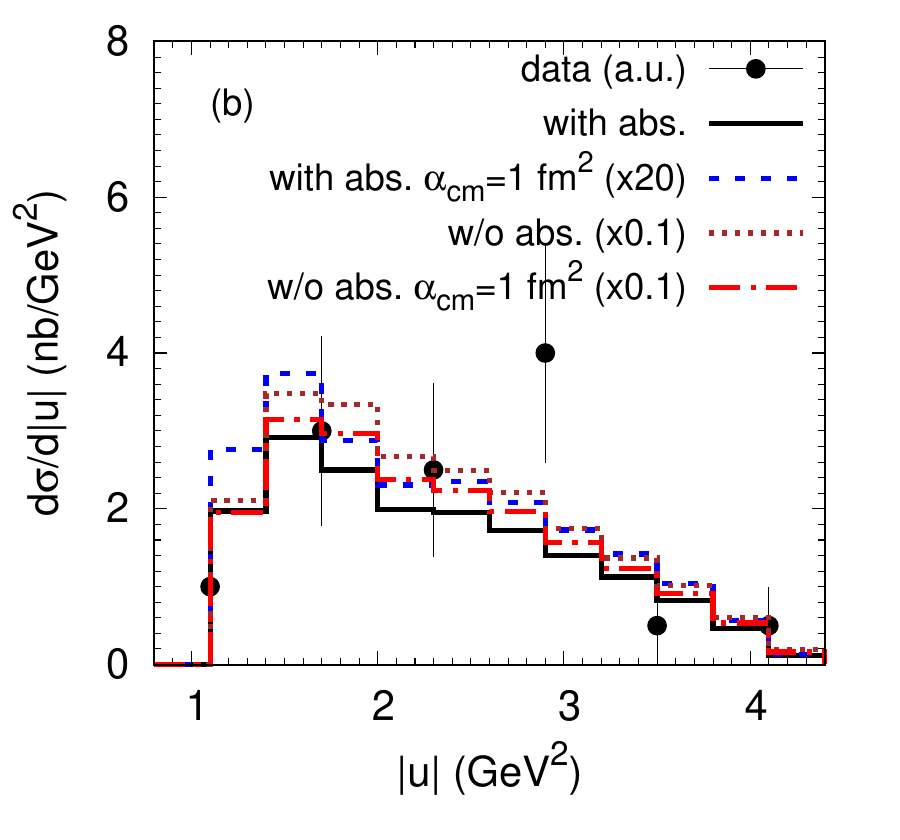}
  \end{center}
  \caption{\label{fig:tu} Distributions of $|t|$ (a) and $|u|$ (b) in the process $^{12}\mbox{C}(p,2pn_s)^{10}\mbox{B}$.
    Calculations using the TISM WF of relative $NN-B$ motion (\ref{psi_20^2}),(\ref{psi_22^2}) with and without absorption are shown by the solid (black)
    and dotted (brown) lines, respectively. Calculations using the lowest HO state WF (\ref{psi_00^2})  with and without absorption are shown by
    the dashed (blue) and dot-dashed (red) lines, respectively. The calculated results are scaled by factors shown in parentheses.
    Experimental data are from Ref.~\cite{Patsyuk:2021fju}.}
\end{figure}

Figure \ref{fig:tu} shows the distributions of the invariants $t=(p_1-p_3)^2$ and $u=(p_1-p_4)^2$. 
The distributions are governed by hard $pp \to pp$ scattering and are sensitive neither 
to the relative WFs of the $NN-B$ motion nor to the absorption. (Note that the insensitivity to the absorption might be partly related
to the fixed kinematics of fast protons in the calculation of the absorption integral, Eq.(\ref{DspaceInt1}), as discussed in the end of sec. \ref{model}.
More precise calculation may change this result but is beyond the scope of our present exploratory study.)
The calculation predicts maxima at $|t|=|u| \simeq 1.5$ GeV$^2$, which are not in the data.
This discrepancy might be, however, attributed to large experimental bins and statistical errors.
\footnote{The calculated $|t|$ and $|u|$ distributions are identical within numerical integration errors ($\sim 10$\%) which
is expected due to the reflection symmetry of experimental setup with respect to the $yz$ plane. 
However, the measured $|t|$ and $|u|$ distributions differ from each other.} 

\begin{figure}
  \begin{center}
  \includegraphics[scale = 0.60]{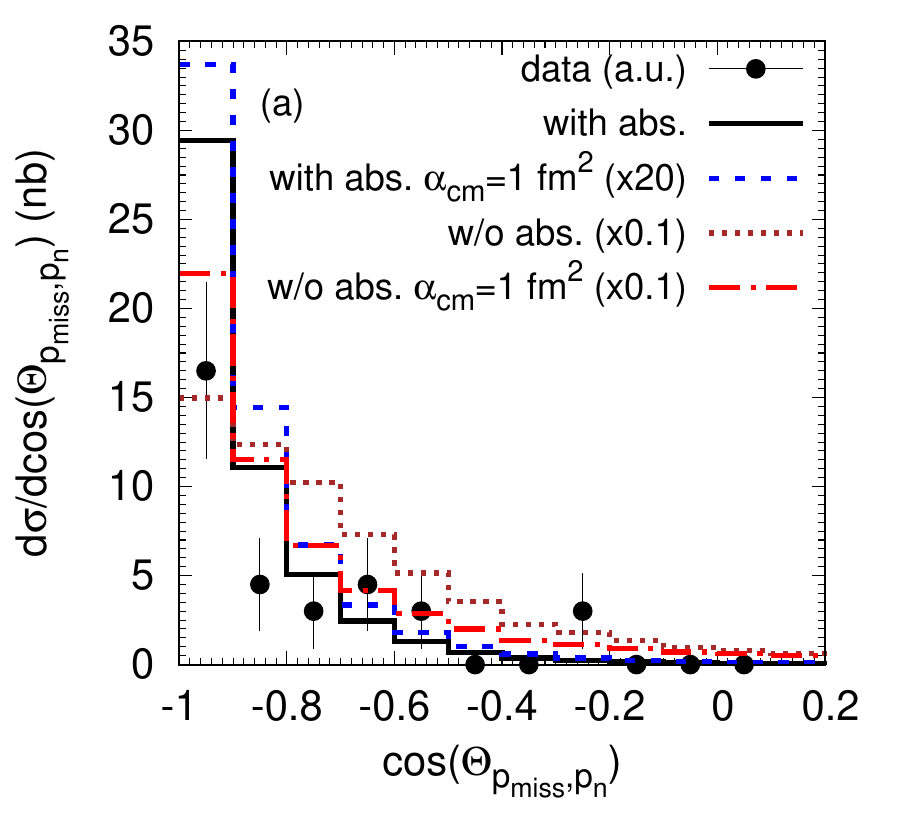}
  \includegraphics[scale = 0.60]{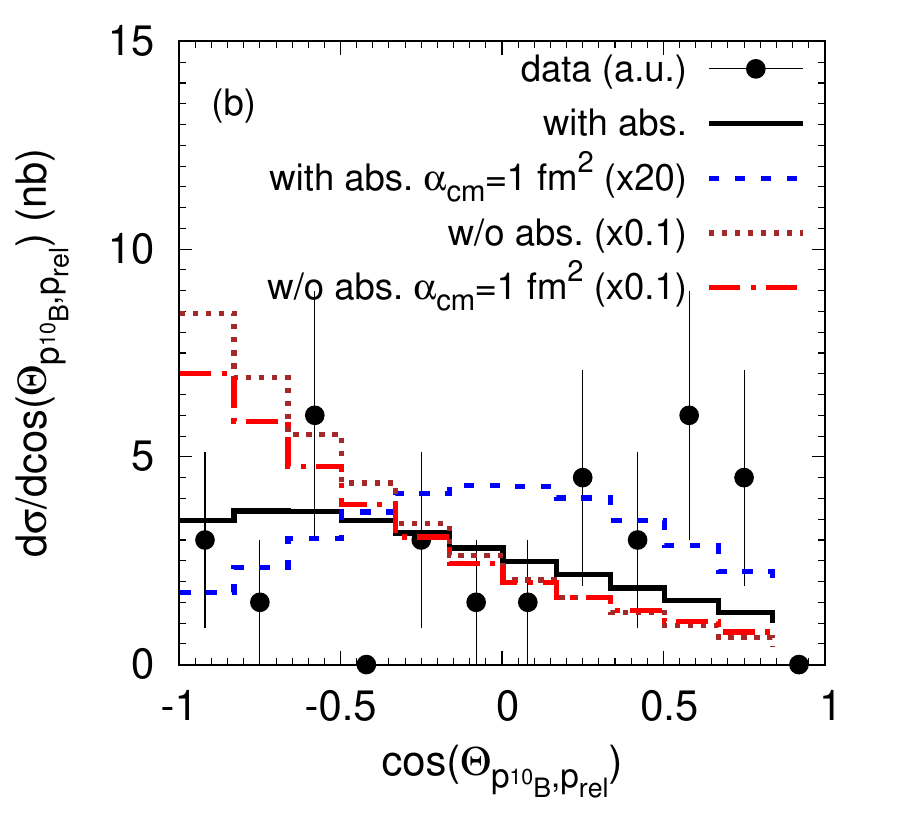}
  \end{center}
  \caption{\label{fig:angles} Distributions of cosine of angle between missing momentum
    and neutron momentum (a) and between $^{10}$B momentum and relative momentum
    $\bvec{p}_{\rm rel} \equiv (\bvec{p}_2-\bvec{p}_5)/2$ (b). Notations are the same as in Fig. \ref{fig:tu}.}
\end{figure}

The distribution of the cosine of angle between $\bvec{p}_{\rm miss} \equiv \bvec{p}_2$ and $\bvec{p}_n \equiv \bvec{p}_5$
in the r.f. of the target nucleus is shown in Fig.~\ref{fig:angles}a.
Without absorption, the phenomenological WF gives a narrower relative $NN-B$ momentum distribution and, therefore, results in a sharper back-to-back correlation
between missing momentum and neutron momentum. Including absorption leads to a somewhat more sharp back-to-back correlation. Similar effect of FSIs was also obtained
in the calculations of Ref.~\cite{Colle:2013nna} for the opening angle distribution of the initial-state protons in the $A(e,e^\prime pp)$ reactions.

Figure \ref{fig:angles}b displays the distribution of the cosine of the angle between the momentum of the outgoing nucleus 
and the relative momentum of nucleons in the $NN$ pair.
The calculations without absorption produce dropping cross sections with decreasing angle and are only weakly sensitive to the WFs
of relative $NN-B$ motion. Including absorption strongly reduces the yield at $180\degree$ due to enhanced absorption of low-energy neutrons.
The phenomenological WF places the $NN$ pair, in-average, closer to the center of the residual nucleus that leads to stronger absorption as compared to the TISM WFs.
As a result, the angle distributions calculated with absorption are sensitive to the WFs.
Note that our calculation with phenomenological WF with absorption gives a weak maximum at $90\degree$, in-line
with the GCF result (see Fig.~4e in Ref.~\cite{Patsyuk:2021fju}).
The BM@N data points seem to indicate no angle dependence. 

\begin{figure}
  \begin{center}
    \includegraphics[scale = 0.60]{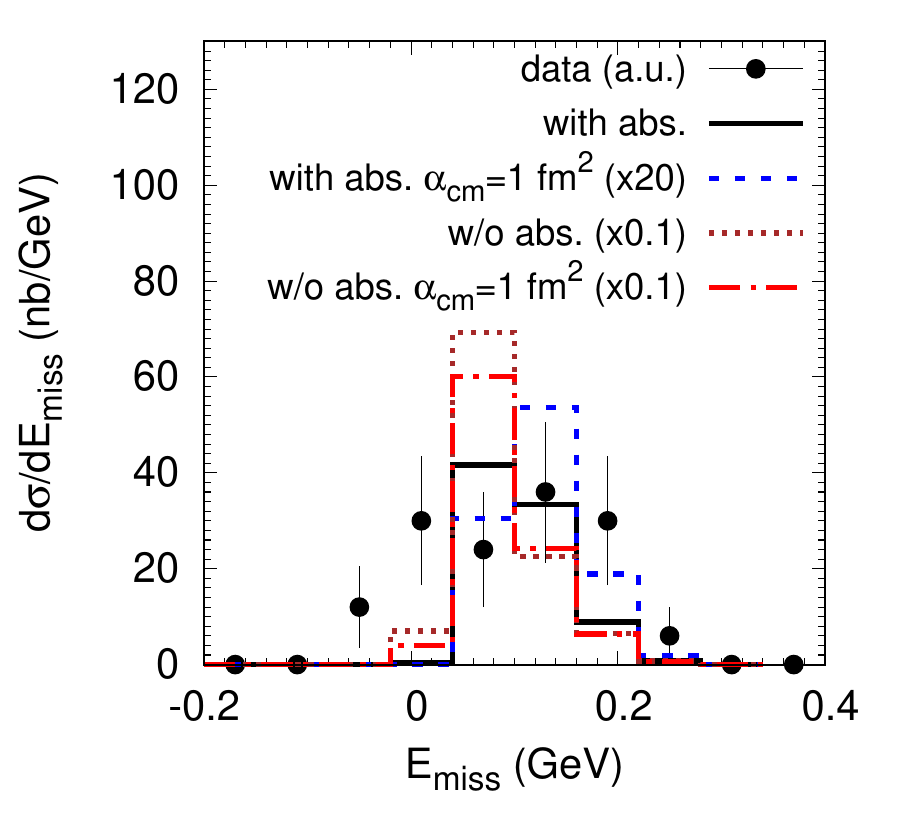}
  \end{center}
  \caption{\label{fig:Emiss} Distribution of missing energy. Notations are the same as in Fig.~\ref{fig:tu}.}
\end{figure}

As is well known, nucleons with high momentum bound in stable nuclei are strongly off mass shell.
This is seen from the distribution of missing energy shown in Fig.~\ref{fig:Emiss}.
Indeed, according to the two-nucleon correlation model \cite{CiofidegliAtti:1991mm},
neglecting c.m. motion of the $NN$ pair, 
the energy of the struck proton can be estimated as follows:
\begin{equation}
  p_2^0 = 2m - B_A + B_{A-2} - \sqrt{\bvec{p}_2^2 + m^2}~, \label{p_2^0}
\end{equation}
where $B_A$ and $B_{A-2}$ are the binding energies of the initial ($^{12}$C) and final ($^{10}$B) nuclei, respectively.
This gives the missing energy
\begin{equation}
  E_{\rm miss} = m - p_2^0 = B_A - B_{A-2} - m + \sqrt{\bvec{p}_2^2 + m^2}~. \label{Emiss}
\end{equation}
Substituting values $B_A=7.68 \times 12$ MeV, $B_{A-2}=6.48 \times 10$ MeV \cite{Audi:2002rp}, and $|\bvec{p}_2|=350-500$ MeV/c
one gets from Eq.(\ref{Emiss}) $E_{\rm miss}=90-152$ MeV in qualitative agreement
with the experimental $E_{\rm miss}$ distribution.
Our numerical results correctly reproduce the centroid position of the measured $E_{\rm miss}$ distribution,
but underestimate the width.
Note that our calculations give a better agreement with experimental
$E_{\rm miss}$ distributions when absorption is included.
This can be again explained by stronger absorption of slow recoil neutron
which corresponds to smaller $E_{\rm miss}$ values.

\begin{figure}
  \vspace{-1cm}
  \begin{center}
  \begin{tabular}{cc}
  \includegraphics[scale = 0.40]{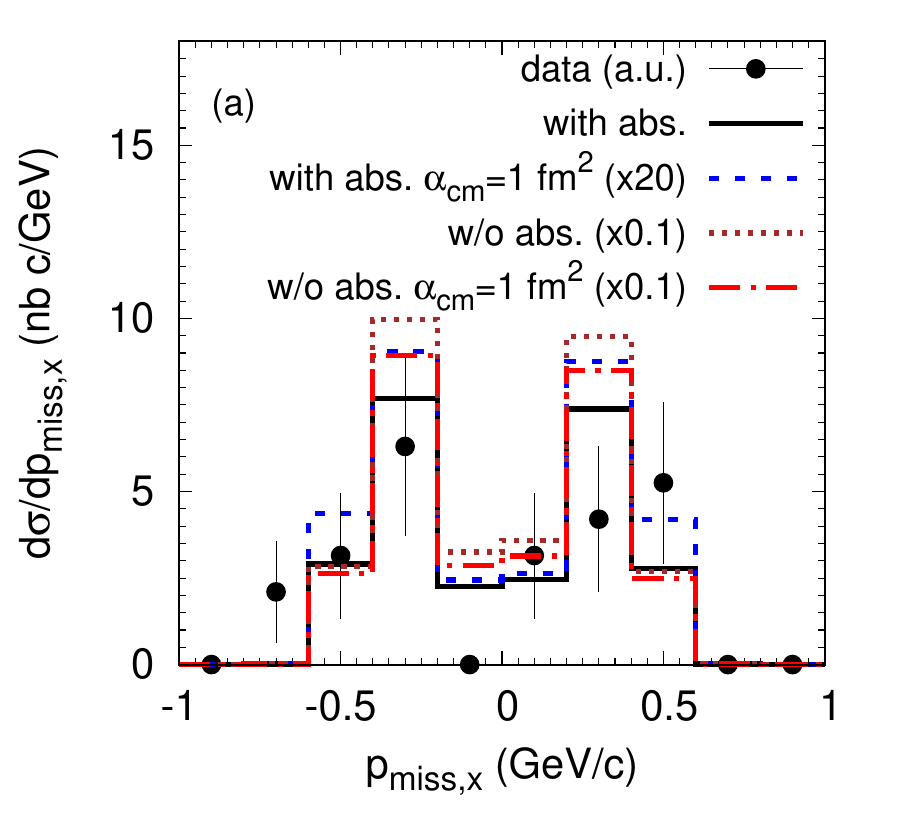} & 
  \includegraphics[scale = 0.40]{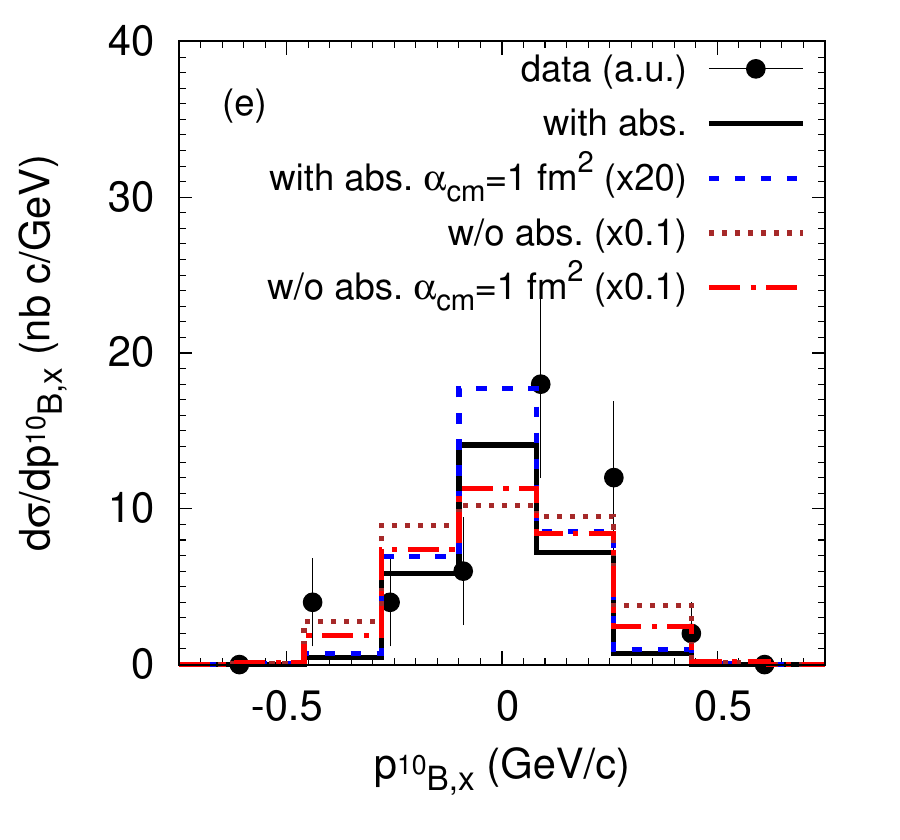} \\
  \includegraphics[scale = 0.40]{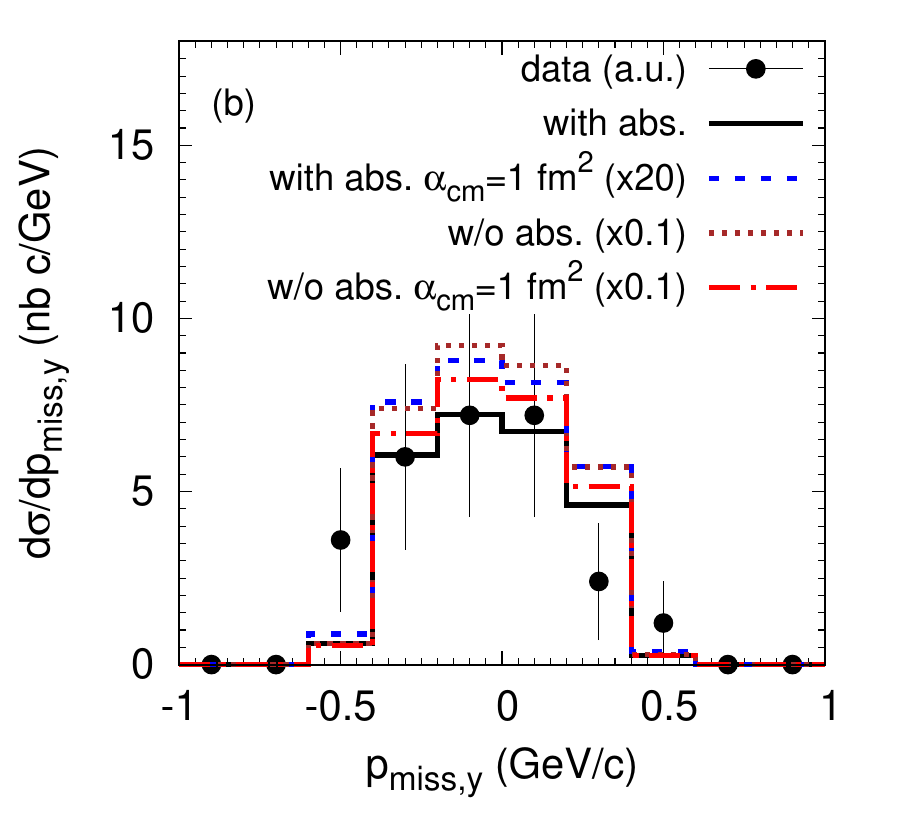} &
  \includegraphics[scale = 0.40]{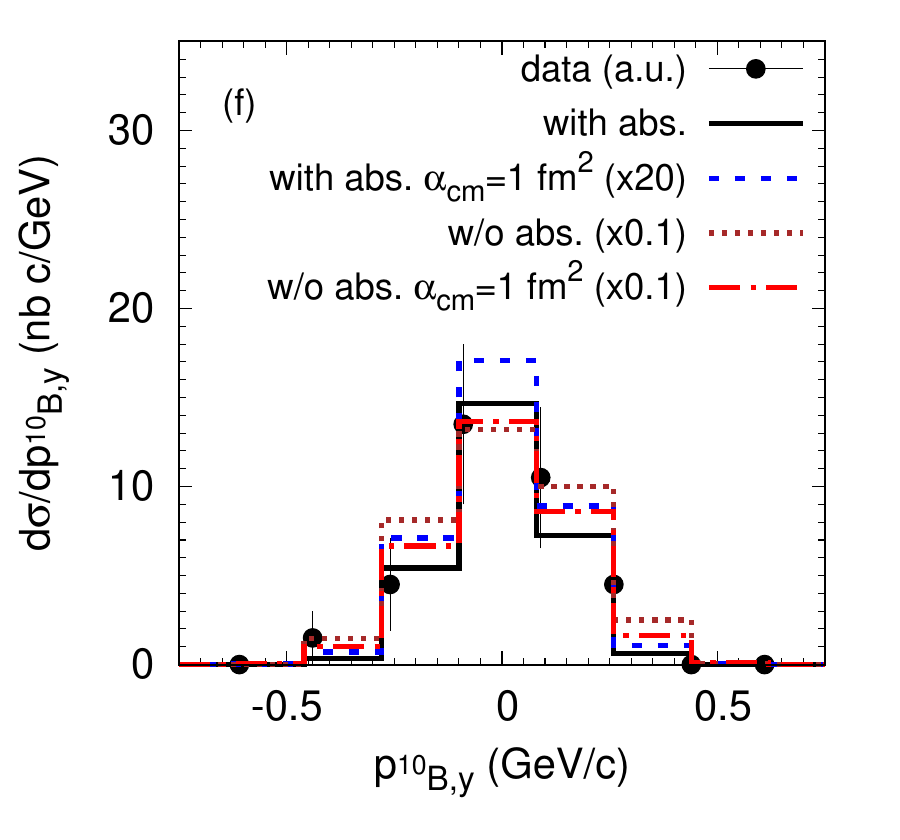} \\   
  \includegraphics[scale = 0.40]{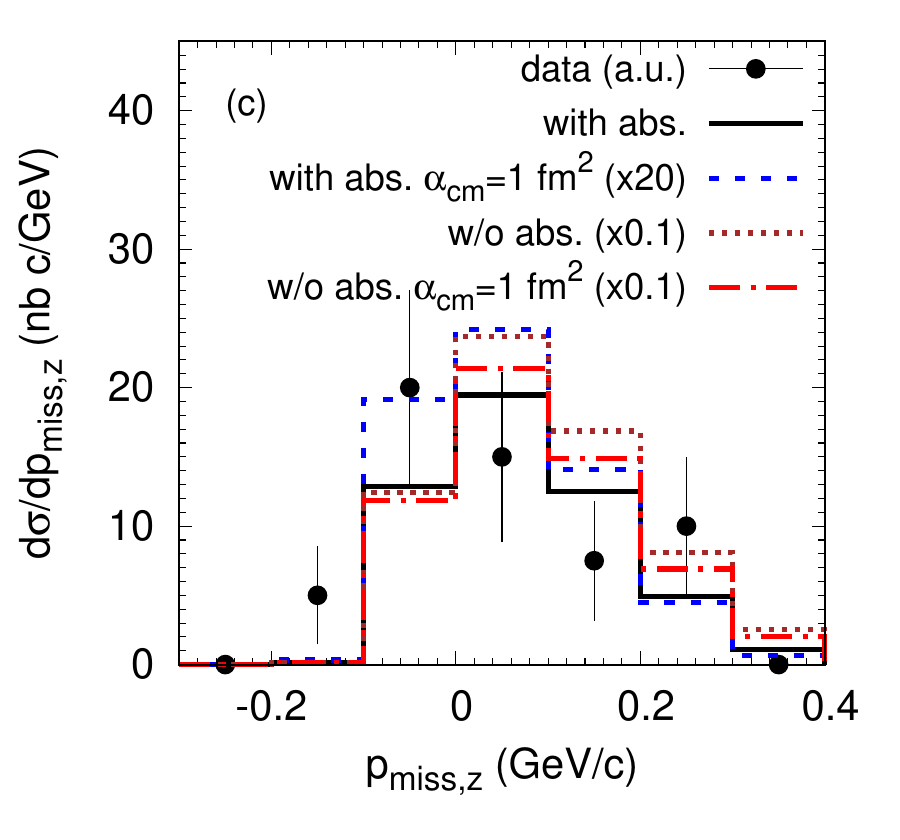} &
  \includegraphics[scale = 0.40]{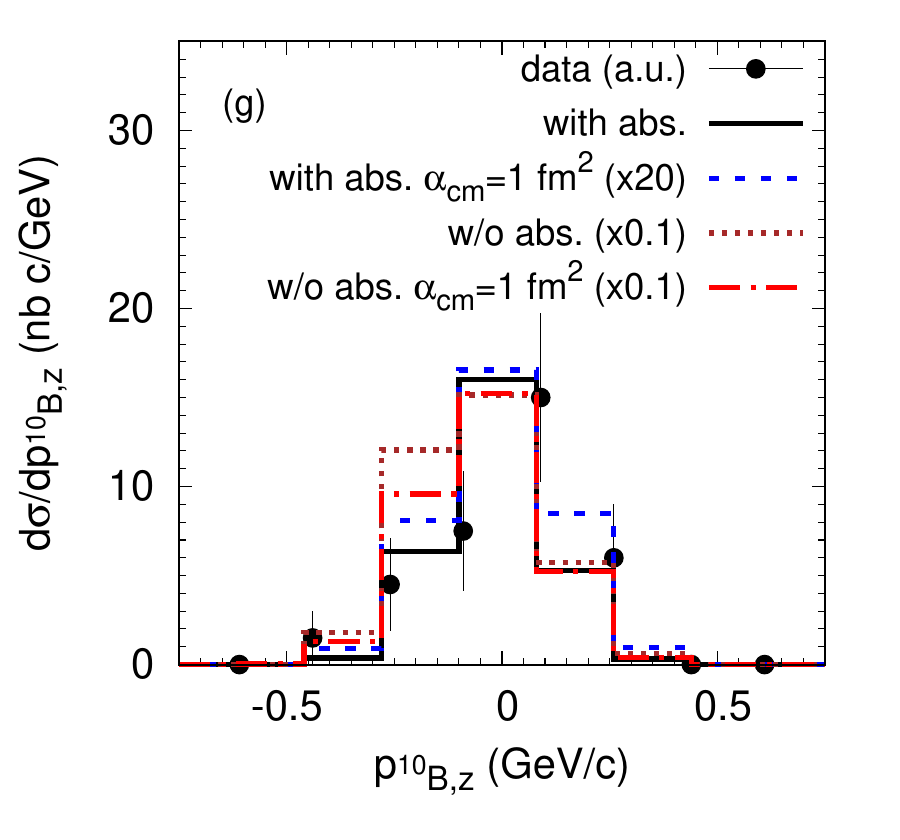} \\ 
  \includegraphics[scale = 0.40]{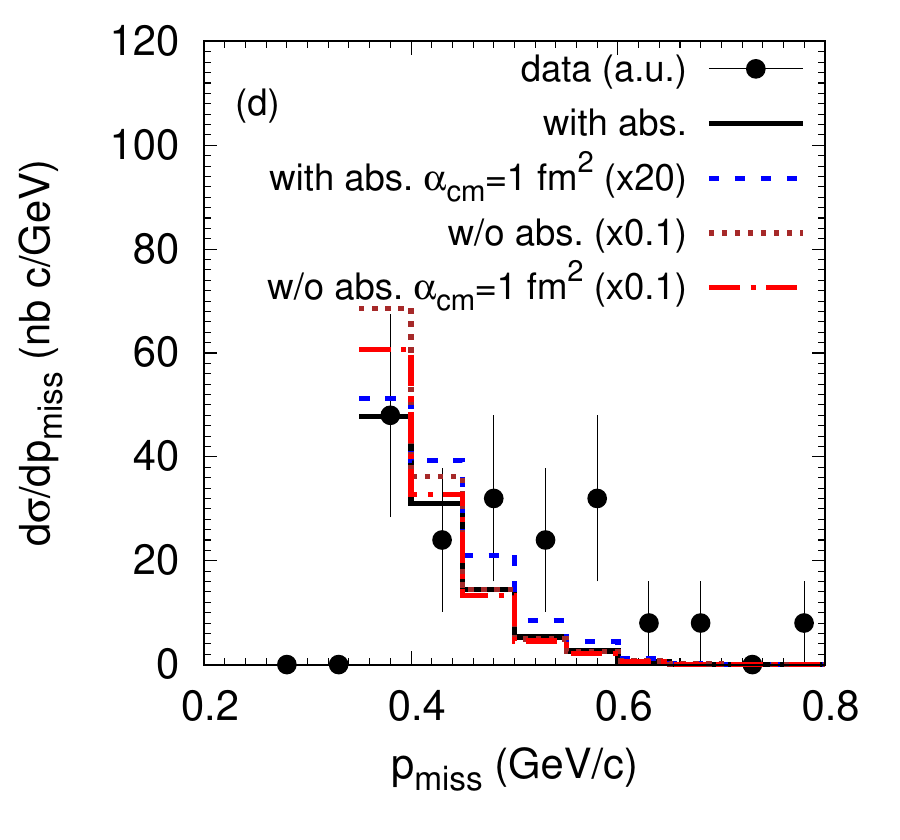} &
  \includegraphics[scale = 0.40]{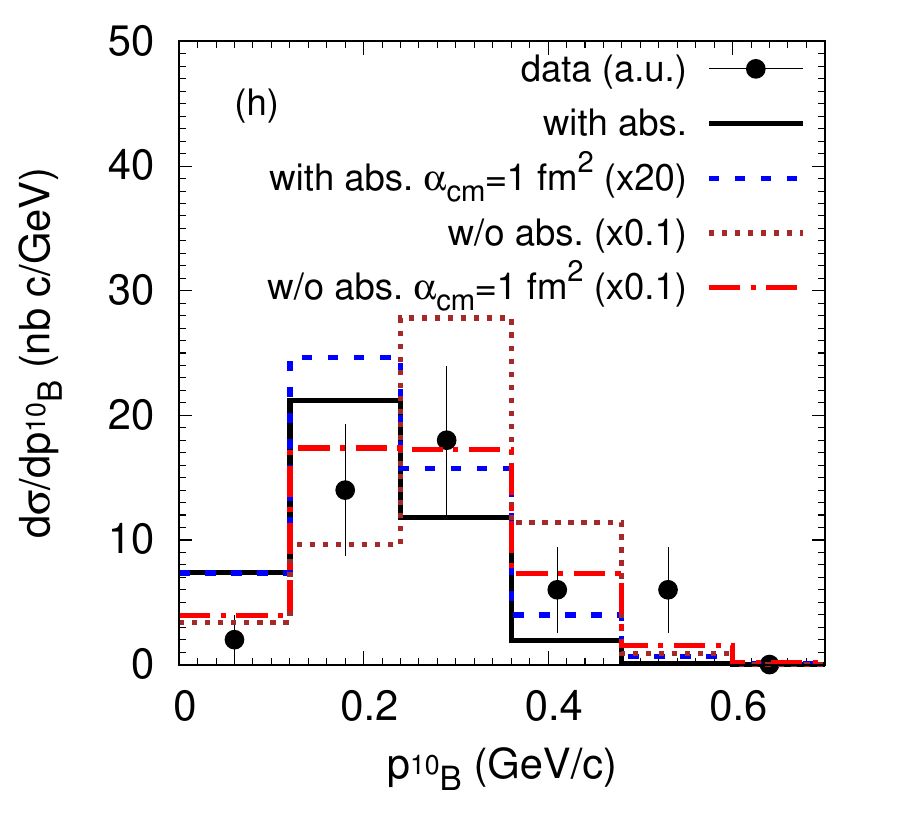} \\
  \end{tabular}
  \end{center}
  \vspace{-1cm}
  \caption{\label{fig:pmiss_pB} Distributions of missing momentum  (a,b,c,d) and of the residual nucleus momentum (e,f,g,h).
    Notations are the same as in Fig.~\ref{fig:tu}.}
\end{figure}

Figure \ref{fig:pmiss_pB}a,b and c show, respectively, the distributions over the $x,y$ and $z$ components of the missing momentum.
Our calculations describe the data quite well irrespective of the presence of absorption and the choice of the WFs.
\footnote{
The calculated $p_{{\rm miss},z}$ distributions are shifted towards positive $p_{{\rm miss},z}$ values.
However, in calculations we applied the $pp \to pp$ differential cross section parameterization
at fixed $\sqrt{s}$ (see sec. \ref{elem}). This does not allow us to explain the shift by 
the $s^{-10}$ scaling of the $pp$ hard elastic cross section at fixed $\Theta_{c.m.}$ \cite{Brodsky:1973kr,Matveev:1973ra}.
The shift is rather caused by the specific angular acceptance of the two-arm spectrometer \cite{Patsyuk:2021fju} configured for
$pp$ scattering at  $\Theta_{c.m.}=90\degree$ for $p_{\rm lab}=4$ GeV/c. Larger $p_{\rm lab}$ corresponding to
negative $p_{{\rm miss},z}$ would decrease the polar scattering angle in the laboratory frame beyond the detector acceptance.
}
Same is true for the $p_{\rm miss}$ distribution (Fig.~\ref{fig:pmiss_pB}d).

Figure \ref{fig:pmiss_pB}e,f and g show, respectively, the distributions over the $x,y$ and $z$ components of the momentum
of the outgoing $^{10}$B. It is expected that these distributions should be most sensitive to the WF of relative
$NN-B$ motion. Indeed, for calculations without absorption, the TISM WFs give broader distributions as compared to those of the phenomenological WF.
Including absorption does not change this conclusion.

Figure \ref{fig:pmiss_pB}h shows the momentum distribution of $^{10}$B. Absorption noticeably influences the spectra at
large values of $p_{^{10}B}$. This is explained by the $p_{\rm miss} > 350$ MeV/c cut. 
As a consequence of momentum conservation in the $^{12}$C rest frame ($\bvec{p}_{\rm miss} + \bvec{p}_n + \bvec{p}_{^{10}B}=0$),
at small $p_{^{10}B}$, this cut selects kinematics with neutrons of larger momenta which experience less absorption, while at large
$p_{^{10}B}$, kinematics with neutrons of smaller momenta, which are suppressed by a stronger absorption, is accepted too. 

\begin{figure}[ht]
  \vspace{-1cm}
  \begin{center}
     \includegraphics[scale = 0.40]{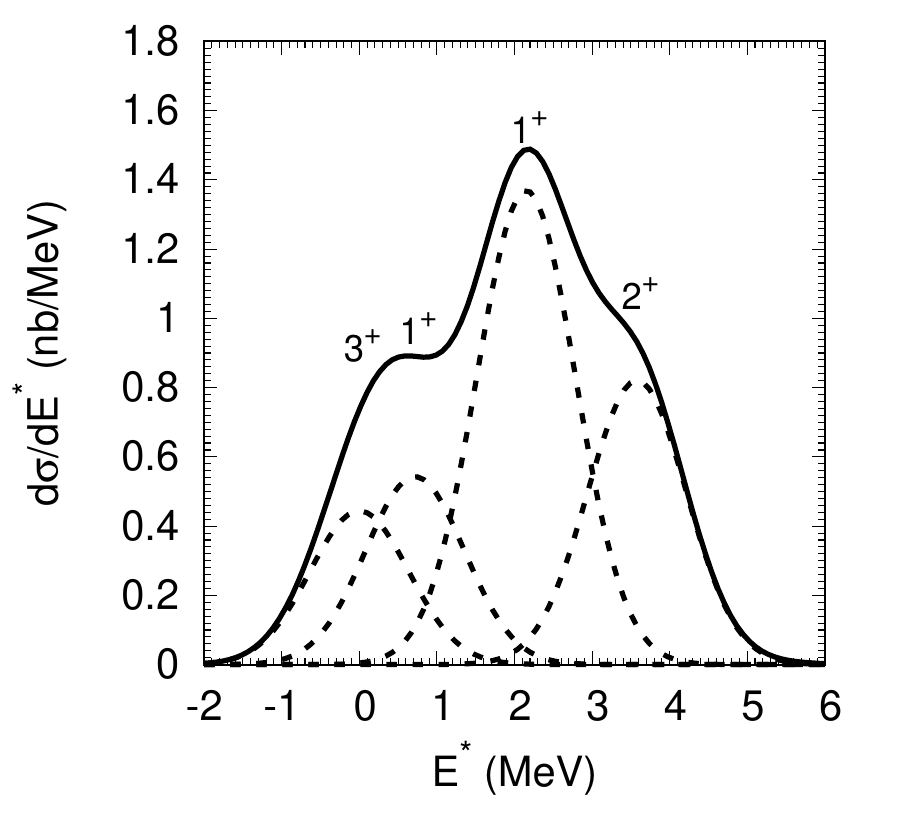}
  \end{center}
  \vspace{-1cm}
  \caption{\label{fig:ExcSpect} Distribution of the residual nucleus $^{10}$B excitation energy (solid line) obtained by
    summing the partial cross sections for
    different energy eigenstates folded for illustration purpose  with the Gaussian distributions of the typical
    experimental resolution FWHM=1.5 MeV
    (dashed lines).}
\end{figure}
Since the TISM calculation directly includes transitions to different internal states of the residual nucleus, it is instructive to
examine partial contributions of various transitions. Fig.~\ref{fig:ExcSpect} shows the excitation energy spectrum of $^{10}$B.
The partial contributions of the $T=0$ states are also shown. ($T=1$ states are included in the total spectrum but not shown since their
contribution is very small.) The spectrum is dominated by the 2.15 MeV $1^+$ state that has 95\% contribution of the $S$-wave.
The 0.717 MeV $1^+$ state has 38\% contribution of the $S$-wave. The $3^+$ ground state and the 3.58 MeV $2^+$ state are pure $D$-wave
ones. Thus, selecting different windows of the excitation  energy it is possible to restrict the partial waves in the relative
$NN-B$ WF. Note that the dominant production of the residual nucleus in the excited (and not ground) state was also found
in the calculations of the $^{12}$C(p,pd)$^{10}$B process in Ref.~\cite{Balashov:1964}.

\begin{figure}[ht]
  \vspace{-1cm}
  \begin{center}
     \includegraphics[scale = 0.40]{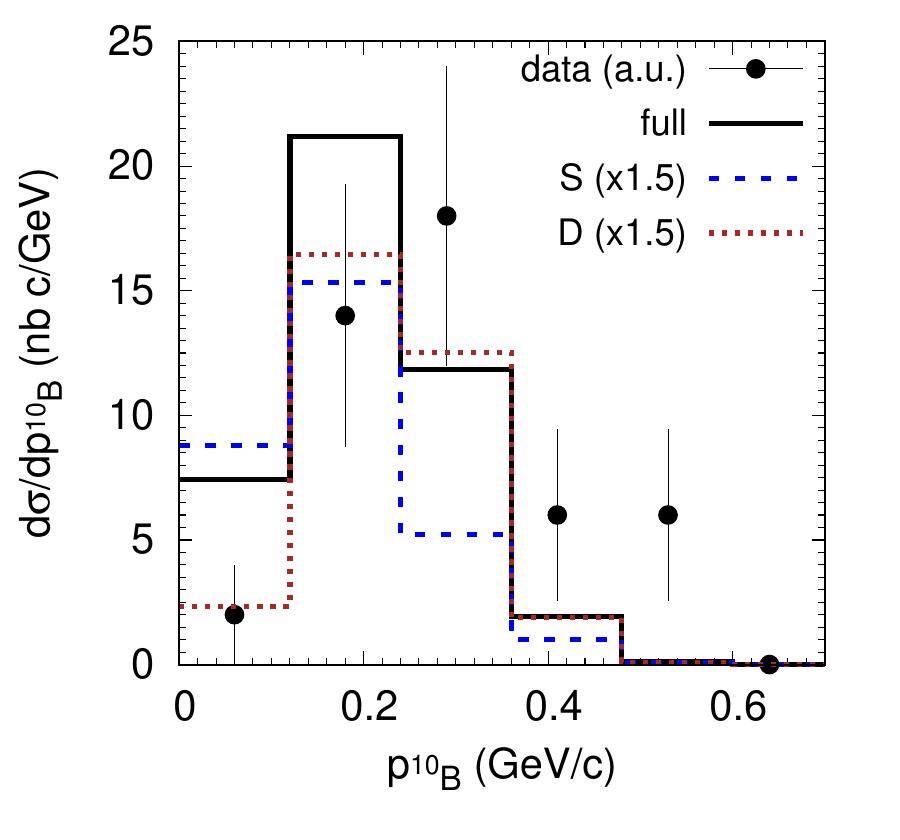}
  \end{center}
  \vspace{-1cm}
  \caption{\label{fig:pB_chk} Distribution of the residual nucleus $^{10}$B momentum. Solid (black) line -- full calculation
    with TISM WFs including absorption. Dashed and dotted lines show, respectively, the partial contributions of transitions to the
    $S$ and $D$ states of $^{10}$B scaled by a factor of 1.5. Experimental data are from Ref.~\cite{Patsyuk:2021fju}.}
\end{figure}
As we see from Table~\ref{tab:FPC}, the orbital angular momentum of the residual nucleus dictates
the angular momentum of the relative $NN-B$ WF and, thus, the residual nucleus momentum distribution
and the absorptive effect of ISI/FSI. In Fig.~\ref{fig:pB_chk}, we examine the momentum distributions of the residual
$^{10}$B nuclei in the $S$- and $D$-wave states. The $D$-wave component has a harder momentum spectrum and seems to agree with data better.
Thus, the present BM@N data may indicate an enhanced contribution of the $D$-wave as compared to the amplitudes of Table~\ref{tab:10B}.   

\begin{table}[t]
  \caption{\label{tab:sig}
    Integrated cross sections (in nb) in the kinematics of the BM@N experiment \cite{Patsyuk:2021fju}.
    Lower line gives the ratio $R=\sigma[^{12}\mbox{C}(p,2pp_s)^{10}\mbox{Be}]/\sigma[^{12}\mbox{C}(p,2pn_s)^{10}\mbox{B}]$  (in \%).
    Results obtained with phenomenological
    relative $NN-B$ WFs (see Eq.(\ref{psi_00^2}))
    are given in parentheses.}
   \begin{center}
    \begin{tabular}{llll}
    \hline
    \hline
        \hspace{4cm}                      & IA \hspace{2cm}      & Abs \hspace{2cm}         & Abs+CEX \\
    \hline
    $^{12}\mbox{C}(p,2pn_s)^{10}\mbox{B}$   & 63.7 (57.1)          & 5.1 (0.31)               & 5.0 (0.29)   \\
    $^{12}\mbox{C}(p,2pp_s)^{10}\mbox{Be}$  & 4.2 (3.3)            & 0.13 (0.0058)            & 0.23 (0.030) \\
       $R$                                & 6.6 (5.8)            & 2.5 (1.8)                & 4.6 (10.4)   \\  
    \hline
    \hline
    \end{tabular}
  \end{center}
\end{table}
We will finally discuss the isospin composition of SRCs. Table~\ref{tab:sig} lists our results for the integrated cross sections of the
two-nucleon knock-out with outgoing $^{10}$B and $^{10}$Be and their ratio $R$. The latter has to be compared with experimental value $R=2/23=(8.7 \pm 6)\%$.
(The total numbers of detected events with $^{10}$Be and $^{10}$B are 2 and 23, respectively, as reported in Ref.~\cite{Patsyuk:2021fju}.
The included statistical error is our estimation.)  

In the IA, the results are not much sensitive to the different WFs of relative $NN-B$ motion. Including absorption reduces cross section by an order of magnitude
in the case of the TISM WFs and by more than two orders -- in the case of phenomenological WFs. Especially strong absorption effect is visible for the channel
with $^{10}$Be for phenomenological WFs.
This is because, on one hand, the $n=0,\Lambda=0$ WF is peaked at $p_B=0$, but in this case the yield is
very strongly suppressed by the $^1S_0$ $NN$ relative WF at $p_{\rm miss} > 350$ MeV/c. On other hand, at finite $p_B$, the yield is suppressed
by strong absorption of low-momentum neutrons.

Including CEX increases the cross section of the $^{10}$Be channel by about 50\% for TISM WFs and five times for phenomenological WFs.
Thus, in the latter case, almost all $^{10}$Be yield is due to the CEX processes. The stronger absorption and CEX for phenomenological WFs
arise from smaller average relative $NN-B$ distances, which force participating nucleons to travel through a region of higher density
of the residual nucleus. Phenomenological WFs provide the best agreement with experiment for the ratio $R$.

\section{Summary}
\label{summary}

We applied the TISM to the hard proton knock-out reactions $^{12}\mbox{C}(p,2pn_s)^{10}\mbox{B}$
and $^{12}\mbox{C}(p,2pp_s)^{10}\mbox{Be}$ for the carbon beam momentum of 48 GeV/c
with an outgoing nucleus in the ground or excited state with excitation energy up to about 6 MeV.
The TISM allowed us to calculate the spectroscopic amplitude for a given quantum states of the
$NN$-pair and residual nucleus including the WF of their relative motion.
The absorptive- and  single-CEX ISI/FSI were taken into account in the eikonal approximation.

We found that absorption reduces the integrated cross section by more than an order of magnitude,
while the CEX processes strongly increase the yield ratio $^{10}\mbox{Be}/^{10}\mbox{B}$.
Absorption and CEX are very sensitive to the WF of the relative $NN-B$ motion.

However, the effect of absorption on the shape of the studied distributions is very moderate,
which was also found in  Ref.~\cite{Uzikov:2021uvh}, where the ISI/FSI effects were estimated
within the framework of a diagrammatic approach for the same reactions.
The strong effect of absorption was observed only for the angular distribution between the momentum of the outgoing nucleus
$^{10}\mbox{B}$ and the relative momentum of the $pn$ pair (Fig.~\ref{fig:angles}b).

The distributions of relative angles, missing momentum, and $^{10}\mbox{B}$ momentum measured by the BM@N Collaboration
\cite{Patsyuk:2021fju} are described quite well by the TISM irrespective of the choice of the WF of the relative $NN-B$ motion
when absorption is taken into account.

The present study is only the first attempt of a detailed comparison of the TISM with SRC data.
In the future, the calculations can certainly be improved, in particular by including the $[431]$ $^{13}P$-configuration of $^{12}$C
and a more accurate description of the ISI/FSI processes.
Availability of more accurate data on SRCs in light nuclei (specific states of the residual nucleus, bigger statistics,
differential cross sections) would be useful to further validate the TISM-based spectroscopic approach.

SRCs may also manifest themselves in a hard knock-out of nuclear clusters. 
A new theoretical analysis of the data \cite{Albrecht:1979zy,Ero:1981zz} on the quasi-elastic knock-out of fast deuterons,
using a similar method of taking into account the effects of ISI/FSI, would be useful.
Of particular interest is the influence of SRCs on cumulative processes \cite{Frankfurt:1979sv},
where our model can also be applied.

\begin{acknowledgments}
  The authors are grateful to Dr. Maria Patsyuk for stimulating discussions and detailed explanations of the BM@N acceptance. 
\end{acknowledgments}

\bibliography{pc12src}

\newpage

\appendix

\section{Calculation of the fractional parentage coefficients of TISM}
\label{FPC}

The FPC of the TISM was calculated using Eq.(\ref{FPCusual}). This equation contains the FPC of the conventional HO model and the cluster coefficient.
In the case of $b=2$, the latter is equal to the Talmi coefficient:
\begin{equation}
  \langle p^2[f_X](\lambda_X\mu_X) {\cal L}S_XT_X | n\Lambda,  2 N_X[f_X](\lambda_X\mu_X)  L_XS_XT_X \rangle
  = \langle 1 1, 1 1 : {\cal L}|11|n\Lambda, N_X  L_X : {\cal L} \rangle~,     \label{TalmiCoeff}
\end{equation}
where the notation of Ref.~\cite{NS} is used in the r.h.s.
The Talmi coefficients are tabulated in Ref.~\cite{NS}. The same tabulation can also be found in Ref.~\cite{Smirnov61}.
\footnote{Note that in Ref.~\cite{Smirnov61} the main oscillator quantum number $n$ is defined so that
the number of oscillator quanta is $2n+l$, which differs from the definition of Ref.~\cite{NS} used in our present work.}
The FPCs for transition to the $^{13}S$ and $^{13}D_I$ states (see Table~\ref{tab:FPC}) are calculated as follows:
\begin{eqnarray}
  \mbox{FPC}(^{13}S) &=& (-1)^2 \times \left(\frac{12}{10}\right)^{2/2}
  \times \left(\begin{array}{c}
    8 \\
    2
  \end{array}\right)^{1/2}
   \times \left(\begin{array}{c}
    12 \\
    2
  \end{array}\right)^{-1/2} \nonumber \\
  && \times \langle p^8[44](04) 000 | p^6[42](22) 010; p^2[2](20) 010 \rangle
     \times \langle 1 1, 1 1 : 0|11|2 0, 0 0 : 0 \rangle  \nonumber \\
  &=&  \frac{6}{5} \times \left(\frac{7 \times 8}{2}\right)^{1/2} \times \left(\frac{11 \times 12}{2}\right)^{-1/2}
  \times \left[\sqrt{\frac{9}{14}}  \times \left(-\sqrt{\frac{16}{54}}\right) \times \sqrt{\frac{1}{2}}\right] \times \sqrt{\frac{1}{2}}  \nonumber \\
  &=& -\sqrt{\frac{8}{275}}~,        \label{FPC_13S}   \\
  \mbox{FPC}(^{13}D_I) &=& (-1)^2 \times \left(\frac{12}{10}\right)^{2/2}
    \times \left(\begin{array}{c}
    8 \\
    2
  \end{array}\right)^{1/2}
   \times \left(\begin{array}{c}
    12 \\
    2
   \end{array}\right)^{-1/2} \nonumber \\
   && \times \langle p^8[44](04) 000 | p^6[42](22) 2_I10; p^2[2](20) 210 \rangle
      \times \langle 1 1, 1 1 : 2|11|2 2, 0 0 : 2 \rangle  \nonumber \\
   &=& \frac{6}{5} \times \left(\frac{7 \times 8}{2}\right)^{1/2} \times \left(\frac{11 \times 12}{2}\right)^{-1/2}
      \times \left[\sqrt{\frac{9}{14}} \times \left(-\sqrt{\frac{3}{54}}\right) \times \sqrt{\frac{1}{2}}\right] \times  \sqrt{\frac{1}{2}}  \nonumber \\
   &=& -\sqrt{\frac{3}{550}}~,    \label{FPC_13D_I}   
\end{eqnarray}
where the FPC of the conventional HO model is given by the product of three factors in the square brackets
corresponding to the weight factor, orbital coefficient, and charge-spin coefficient
tabulated in Ref.~\cite{Elliott53}. The FPC for transition to the $^{13}D_{II}$ state is obtained by replacing the orbital coefficient
$-\sqrt{\frac{3}{54}} \to -\sqrt{\frac{35}{54}}$ in Eq.(\ref{FPC_13D_I}).
The FPCs for transitions to the $^{31}S$, $^{31}D_I$ and  $^{31}D_{II}$ states are obtained by
replacing the charge-spin coefficient $\sqrt{\frac{1}{2}} \to -\sqrt{\frac{1}{2}}$ in the FPCs for transition to
the $^{13}S$, $^{13}D_I$ and  $^{13}D_{II}$ states, respectively.

\end{document}